\documentclass[proceedings]{JHEP37}
\usepackage{eurosym}
\usepackage{amssymb}
\usepackage{amsfonts}
\usepackage{amsmath,epsfig}
\usepackage{graphicx}

\newbox\mybox

\newcommand\fverb{\setbox\mybox=\hbox\bgroup\verb}
\newcommand\fverbdo{\egroup\medskip\noindent\fbox{\unhbox\mybox}\ }
\newcommand\fverbit{\egroup\item[\fbox{\unhbox\mybox}]}
\conference{Complex BPS Skyrmions with real energy}
\abstract{We propose and investigate several complex versions of extensions and restrictions of the Skyrme model with a well-defined
Bogomolny-Prasad-Sommerfield (BPS) limit. The models studied possess complex kink, anti-kink, semi-kink, massless and purely imaginary compacton BPS solutions that all have real energies. The 
reality of the energies for a particular solution is guaranteed when a modified antilinear $\mathcal{CPT}$-symmetry maps the Hamiltonian functional to its parity time-reversed complex conjugate
 and the solution field to 
itself or a new field with degenerate energy. In addition to the known BPS Skyrmion configurations we find new types that we refer to as step, cusp, shell, and purely imaginary compacton solutions.}

\title{Complex BPS Skyrmions with real energy}
\author{Francisco Correa$^\circ$, Andreas Fring$^\bullet$ and Takanobu Taira$%
^\bullet$ \\
$\bullet$ Department of Mathematics, City, University of London,\\
$\,\,$ Northampton Square, London EC1V 0HB, UK \\
$\circ$ Instituto de Ciencias F{\'{\i}}sicas y Matem{\'{a}}ticas,
Universidad Austral de Chile, \\
$\,\,$ Casilla 567, Valdivia, Chile\\
E-mail: francisco.correa@uach.cl, a.fring@city.ac.uk,
takanobu.taira@city.ac.uk}

\input{tcilatex}
\begin{document}

\section{Introduction}

The Skyrme model \cite{skyrme1962unified} has been introduced as a potential
candidate for a low energy effective field theoretical description of a
strongly interacting matter theory, i.e. Quantum Chromodynamics, more than
fifty years ago. It took a fairly long time to demonstrate that the model
could indeed arise as such type of low energy effective theory in a limit
for which the number of quark colours is taken to be very large \cite%
{witten1983current1,witten1983current2}. The Skyrme model is perfectly
tailored to the nonperturbative nature of that energy regime and
successfully describes various key characteristics of atomic nuclei. The
topological soliton solutions of the model, the Skyrmions, are identified as
Baryons with integer topological charges $B$ being elements in the third
homotopy group for the $SU(2)$-group valued fields, $B\in \mathbb{Z\simeq
\pi }_{3}(SU(2))$. The field excitations around a trivial vacuum are
identified as pions \cite%
{adkins1984skyrme,battye2006skyrmions,harland2014topological} and there
exist also variants of the model that include $\rho $, $\omega $ and $A_{1}$
vector mesons \cite{meissner1986skyrmions,bando1985rho}. Early on in the
exploration of the model it was also noticed that Skyrmions allow for a
fermionic interpretation \cite{finkelstein1968connection} and that one may
formulate gauge theoretical versions of them \cite{faddeev1976some}. Atiyah
and Manton established the remarkable fact that static Skyrmion solutions in 
$\mathbb{R}^{3}$ can be approximated well by holonomies of $SU(2)$
Yang-Mills instantons in $\mathbb{R}^{4}$ \cite%
{atiyah1989skyrmions,sutcliffe2010skyrmions}.

Despite the success of the model on qualitative and conceptual aspects, it
is still way off on a quantitative level when comparing numerical solutions
to experimental measurements \cite{adkins1983static}, as most quantities
differ by a fair amount, such as for instance the magnetic moments for the
protons and neutrons which are too small by about 30\%. The Euler-Lagrange
equation associated to the Skyrme model is a complicated nonlinear wave
equation for which various solutions have been obtained numerically for
small and large Baryon numbers \cite%
{battye2001solitonic,battye2002skyrmions,khomotopy,battye2006skyrmions,feist2013skyrmions,gillard2015skyrmions,manton2004topo}%
. The energies for all these solutions show that the binding energy, that is
the energy required to separate a multi-Skyrmion solution into single
Skyrmions normalized by the Baryon number, is far too large when compared to
what is expected from experiments. Motivated by trying to address this
discrepancy, different variants of the original Skyrme model have been
explored. Especially promising are versions of the model with a well-defined
Bogomolny-Prasad-Sommerfield (BPS) \cite%
{bogomol1976stability,prasad1975exact} limit as originally proposed in \cite%
{adam2010skyrme}.

These models exhibit a number of very appealing features: Firstly, they
allow for the construction of elegant exact analytical solutions in form of
topological solitons that satisfy the Bogomolny bounds. Secondly they
reproduce the linear relation between the binding energies and the baryon
number for small and large values. Thirdly, and most importantly, they\
resolve the issue of the discrepancy of the large binding energies in the
original Skyrme model. In fact, in the BPS versions of the model the binding
energies are zero and one may adopt the view that quantum corrections will
only introduce small variations, hence producing the expected smaller values
for the binding energies. Taking corrections from collective coordinate
quantization of spin and isospin, the electrostatic Coulomb energies, and
small explicit breaking of the isospin symmetry into account lead to a very
good agreement between theory and experimental values for the binding energy
as shown in \cite{adam2013bogomol,bonen2010,bonen2012}. For a recent review
on these type of BPS Skyrme models see \cite{adam2017skyrme}.

Motivated by the success of the BPS versions of the original Skyrme model,
we explore here further possible variants that include complex non-Hermitian
versions of these models. We demonstrate that some of their static solutions
have real topological energies despite being complex and thus these
solutions may also be associated to well-defined physical objects.

In order to overcome the so-called auxiliary field problem and emergence of
fourth-order time derivatives when introducing supersymmetry, complex
versions of the Skyrme model were previously studied \cite%
{gudnasonnitta1,gudnasonnitta2}, by taking the fields to be valued in a
complexification of $SU(N)$, i.e. $SL(N,\mathbb{C})$. Besides trying to
overcome the above mentioned problems, these studies were guided by the fact
that the introduction of supersymmetry into the Skyrme model is almost
inevitably forcing the introduction of complex structures as the underlying
manifolds need to be of K\"{a}hler type for this purpose. However, complex
solutions and their reality conditions were not considered previously.

As argued in \cite{fring2020BPS} the reality of the energy for some scalar
field solutions $\mathcal{\phi }_{i}$, $i=1,2,\ldots $, to the BPS equations
or the equations of motion is guaranteed when the following three conditions
are met:

\begin{description}
\item[(i)] There exists a modified $\mathcal{CPT}$-symmetry that maps the
Hamiltonian functional to its parity time-reversed complex conjugate 
\begin{equation}
\mathcal{CPT}:\mathcal{H[}\phi (x_{\mu })\mathcal{]\rightarrow H}^{\dagger }%
\mathcal{[}\phi (-x_{\mu })\mathcal{]},  \label{cpt1}
\end{equation}%
with $\mathcal{CPT}^{2}=\mathbb{I}$. Here $\mathcal{CPT}$-symmetry is not to
be taken literally as a simultaneous parity, time reversal and charge
conjugation, but be understood simply as an antilinear map of any kind in
the sense described by Wigner in \cite{EW}.

\item[(ii)] Two solutions $\mathcal{\phi }_{i}$ and $\mathcal{\phi }_{j}$,
not necessarily distinct, are related to each other by the modified $%
\mathcal{CPT}$-symmetry as 
\begin{equation}
\mathcal{CPT}:\mathcal{\phi }_{i}(x_{\mu })\mathcal{\rightarrow \phi }%
_{j}(-x_{\mu }).  \label{cpt2}
\end{equation}

\item[(iii)] The energies $E[\mathcal{\phi }]$ of the two solutions $%
\mathcal{\phi }_{i}$ and $\mathcal{\phi }_{j}$ are degenerate%
\begin{equation}
E[\mathcal{\phi }_{i}]=E[\mathcal{\phi }_{j}].  \label{cpt3}
\end{equation}%
Evidently when $\mathcal{\phi }_{i}=\mathcal{\phi }_{j}$ this condition
holds trivially and the energy is automatically guaranteed to be real. When $%
\mathcal{\phi }_{i}\neq \mathcal{\phi }_{j}$ we must ensure that the
energies are degenerate to reach the same conclusion. As we shall see below
this is often a consequence of some symmetries in some coupling or
integration constants or by the fact that energies for solutions of the
self-dual and anti-self-dual BPS equations are identical as argued in \cite%
{fring2020BPS}.
\end{description}

For a more detailed reasoning on why these conditions and further examples
we refer the reader to \cite{fring2020BPS} and references therein. We will
present examples below for models with solutions satisfying all three
conditions so that their energies are real, but we shall also explore the
broken $\mathcal{CPT}$-regime by presenting counter examples for solutions
with complex energies for which either or both conditions (ii) and (iii) do
not hold.

Our manuscript is organized as follows: In section 2 we recall a general
Lagrangian density that encompasses a whole set of extensions and
restrictions of the standard version of the Skyrme model. In sections 3 we
discuss a complex, albeit pseudo Hermitian, version of the Skyrme model that
possess new types of solutions that satisfy all three conditions (\ref{cpt1}%
)-(\ref{cpt3}) and have therefore real energies. In section 4 we discuss a
version of the BPS Skyrme model with a potential that leads to solutions
that behave asymptotically as kinks but with finite values a zero and also
massless solutions with zero energy. In section 5 we discuss a version of
the model involving a whole ray of Bender-Boettcher type potentials that
possess fractional compacton and semi-kink solutions with real energies. In
section 6 we explore the broken $\mathcal{CPT}$-regime by discussing a model
for which either condition (ii) and/or condition (iii) are not satisfied.
Section 7 contains a discussion of a Skyrmion submodel with complex
semi-kink and soliton-like solutions. Our conclusions are stated in section
8.

\section{The Skyrme model - extensions and restrictions}

To establish our notations and conventions we briefly recall some key
aspects and definitions of the Skyrme model. Largely following \cite%
{adam2010skyrme,adam2017skyrme}, we consider an extended version of the
standard Skyrme model described by variants of a Lagrangian density of the
general form%
\begin{equation}
\mathcal{L=\tilde{L}}_{0}+\mathcal{L}_{2}+\mathcal{L}_{4}+\mathcal{L}_{6}+%
\mathcal{L}_{0},  \label{L}
\end{equation}%
where the different terms are defined as%
\begin{equation}
\mathcal{L}_{2}:=-\frac{f_{\pi }^{2}}{2}\limfunc{Tr}\left( L_{\mu }L^{\mu
}\right) ,~~\mathcal{L}_{4}:=\frac{1}{16e^{2}}\limfunc{Tr}\left( [L_{\mu
},L_{\nu }]^{2}\right) ,~~\mathcal{L}_{6}:=-\lambda ^{2}N_{0}^{2}B_{\mu
}B^{\mu },~~\mathcal{L}_{0}:=-\mu ^{2}V,
\end{equation}%
with Lie algebraic currents in form of right Maurer Cartan forms,
topological current and $SU(2)$-group valued Skyrme fields%
\begin{equation}
L_{\mu }:=U^{\dagger }\partial _{\mu }U,~~~~~B^{\mu }:=\frac{1}{N_{0}}%
\varepsilon ^{\mu \nu \rho \tau }\limfunc{Tr}\left( L_{\nu }L_{\rho }L_{\tau
}\right) ,~~~~~U:=e^{i\zeta (\sigma \cdot \vec{n})},
\end{equation}%
respectively. Here $f_{\pi }$ can be interpreted as the pion decay constant
and the dimensionless constant $e$ is referred to as the Skyrme parameter.
As is well known, these parameters can be scaled away, so that we may set
them both to $1$ in what follows. Moreover, we denote by $\sigma $ the
standard Pauli matrices and take the three component unit vector to be of
the form $\vec{n}=(\sin \Theta \cos \Phi ,\sin \Theta \sin \Phi ,\cos \Theta
)$ rather than the rational map or stereographic projection often used
instead in this context, see e.g. \cite{hrational}. Our space-time metric $g$
is taken to be $\limfunc{diag}g=(1,-1,-1,-1)$. The normalization constant $%
N_{0}$ is chosen in such a way that the Baryon number $B=\int B_{0}d^{3}x\in 
\mathbb{Z}$ becomes an integer as it should be for a two flavour theory to
guarantee that Baryons with an even and odd number of quarks are Bosons and
Fermions, respectively. See for instance \cite{witten1983current2} for a
more detailed reasoning on this issue. For a standard static compacton
solution the normalization constant is usually taken to be $N_{0}=24\pi ^{2}$%
.

Dropping and decomposing terms or further specifying the potential in the
general Lagrangian $\mathcal{L}$ gives rise to different versions of the
model. The original Skyrme model \cite{skyrme1962unified} is comprised of
the sum of the sigma model term $\mathcal{L}_{2}$ and the Skyrme term $%
\mathcal{L}_{4}$ with occasionally the potential term $\mathcal{\tilde{L}}%
_{0}$ added which is of the same functional form as $\mathcal{L}_{0}$. The
BPS version of the model introduced in \cite{adam2010skyrme} consists of the
sum of $\mathcal{L}_{6}$, that mimics the interactions generated by the
vector mesons, and the potential term $\mathcal{L}_{0}$.

Consistent submodels may be obtained by further decomposing terms in $%
\mathcal{L}$. With our choice of the parameterization for the $SU(2)$-group
valued element $U$ the various parts of the Lagrangian take on the following
forms: For reasons that will become clear below, we decompose the sigma
model and the Skyrme term as 
\begin{equation}
\mathcal{L}_{2}=\mathcal{L}_{2}^{(1)}+\mathcal{L}_{2}^{(2)},~~\ \ \text{%
and~~\ \ ~}\mathcal{L}_{4}=\mathcal{L}_{4}^{(1)}+\mathcal{L}_{4}^{(2)},
\end{equation}%
with%
\begin{eqnarray}
\mathcal{L}_{2}^{(1)} &=&\sin ^{2}\zeta \left( \Theta _{\mu }\Theta ^{\mu
}+\Phi _{\mu }\Phi ^{\mu }\sin ^{2}\Theta \right) ,~~\  \\
\mathcal{L}_{2}^{(2)} &=&\zeta _{\mu }\zeta ^{\mu }, \\
\mathcal{L}_{4}^{(1)} &=&\sin ^{2}\zeta \left[ \Theta _{\mu }\zeta ^{\mu
}\Theta _{\nu }\zeta ^{\nu }-\Theta _{\mu }\Theta ^{\mu }\zeta _{\nu }\zeta
^{\nu }+\sin ^{2}\Theta \left( \Phi _{\mu }\zeta ^{\mu }\Phi _{\nu }\zeta
^{\nu }-\Phi _{\mu }\Phi ^{\mu }\zeta _{\nu }\zeta ^{\nu }\right) \right] ,
\\
\mathcal{L}_{4}^{(2)} &=&\sin ^{4}\zeta \sin ^{2}\Theta \left( \Theta _{\mu
}\Phi ^{\mu }\Theta _{\nu }\Phi ^{\nu }-\Theta _{\mu }\Theta ^{\mu }\Phi
_{\nu }\Phi ^{\nu }\right) .
\end{eqnarray}%
The extended part computes with%
\begin{equation}
B^{\mu }=\frac{1}{2N_{0}}\sin ^{2}\zeta \sin \Theta \,\mathcal{B}^{\mu
},~~~\ \ \ \mathcal{B}^{\mu }:=\varepsilon ^{\mu \nu \rho \tau }\zeta _{\nu
}\Theta _{\rho }\Phi _{\tau }
\end{equation}%
to%
\begin{equation}
\mathcal{L}_{6}=-\frac{\lambda ^{2}}{4}\sin ^{4}\zeta \sin ^{2}\Theta 
\mathcal{B}_{\mu }\mathcal{B}^{\mu }=\frac{\lambda ^{2}}{4}\sin ^{4}\zeta
\sin ^{2}\Theta \left[ \varphi _{0}^{a}\mathcal{Q}_{a}^{i}\varphi _{0}^{b}%
\mathcal{Q}_{b}^{i}-\mathcal{B}_{0}\mathcal{B}_{0}\right] ,
\end{equation}%
where $\mathcal{Q}_{a}^{i}:=\frac{1}{2}\varepsilon _{abc}\varepsilon
^{ijk}\varphi _{j}^{b}\varphi _{k}^{c}$, $\varphi :=(\zeta ,\Theta ,\Phi )$
and $a,b,c,i,j,k\in \{1,2,3\}$.

Finally, the pion mass term in the standard BPS version of the model (BPSS) $%
\mathcal{L}_{0}^{BPSS}=-\mu ^{2}V$ is taken to involve the potential $V=%
\frac{1}{2}\limfunc{Tr}\left( \mathbb{I-}U\right) $ $=1-\cos \zeta $, but we
will allow here other forms of the potential as well. Further extensions,
including for instance a sextic derivative term \cite{gudnason2018expl} or
multiplying the terms with field dependent coupling constants \cite%
{adam2020die} have also been studied.

In what follows we shall investigate different combinations of various
complex extended or deformed versions of different parts of this model
related to the form of $\mathcal{L}$ in (\ref{L}).

\section{Pseudo Hermitian variants of Skyrme models}

In this section our first guiding principle is to identify a $\mathcal{CPT}$%
-symmetry in a Hermitian Hamiltonian and extend the model by deforming or
adding complex terms to convert it into a non-Hermitian Hamiltonian that
still respects this symmetry. Subsequently we try to identify a pseudo
Hermitian counterpart in a similar fashion as what is by now standard for
non-Hermitian quantum mechanical systems \cite{PTbook,Alirev}. For BPS
systems in 1+1 dimensions this approach was recently applied successfully in 
\cite{fring2020BPS}. We shall now demonstrate that it can also be applied to
3+1 dimensional theories with complex topological solutions.

\subsection{Complex boosted BPS Skyrme models}

We start with the standard BPS Skyrme model consisting of $\mathcal{L}_{6}+%
\mathcal{L}_{0}^{BPSS}$ by noting that it remains invariant under the
antilinear $\mathcal{CPT}$-transformation: $\zeta \rightarrow -\zeta $, $%
\imath \rightarrow -\imath \,$. Thus we may introduce a complex shift in $%
\zeta \rightarrow \zeta +\imath \kappa ~$with $\kappa \in \mathbb{R}$
without breaking that symmetry. We denote here and in what follows the
imaginary unit as $\imath :=\sqrt{-1}$ to distinguish it from indices $i$.
Choosing $\kappa =$ $-\func{arctanh}\epsilon $ with $\epsilon \in \mathbb{R}$
and using the identities $\sqrt{1-\epsilon ^{2}}\sin \left( \zeta -\imath 
\func{arctanh}\epsilon \right) =\sin \zeta -\imath \epsilon \cos \zeta $, $%
\sqrt{1-\epsilon ^{2}}\cos \left( \zeta -\imath \func{arctanh}\epsilon
\right) =\cos \zeta +\imath \epsilon \sin \zeta $, we obtain a $\mathcal{CPT}
$-symmetrically extended BPS Skyrme model of the form 
\begin{equation}
\mathcal{L}_{\text{b}}=-\frac{\lambda ^{2}}{4}\left( \sin \zeta -\imath
\epsilon \cos \zeta \right) ^{4}\sin ^{2}\Theta \mathcal{B}_{\mu }\mathcal{B}%
^{\mu }-\mu ^{2}\left( \sqrt{1-\epsilon ^{2}}-\cos \zeta -\imath \epsilon
\sin \zeta \right) ,  \label{Lb}
\end{equation}%
after re-scaling the coupling constants as $\lambda \rightarrow \lambda
(1-\epsilon ^{2})$, $\mu \rightarrow \mu (1-\epsilon ^{2})^{1/4}$. By
design, for vanishing $\epsilon $ the model reduces to the standard BPS
Skyrme model $\lim_{\epsilon \rightarrow 0}\mathcal{L}_{\text{b}}=\mathcal{L}%
_{6}+\mathcal{L}_{0}^{BPSS}$ as introduced and discussed in \cite%
{adam2010skyrme}. We shall now demonstrate that the energies for the
topological solutions to the equations of motion resulting from $\mathcal{L}%
_{\text{b}}$ and its corresponding Hermitian counterpart are identical and
real.

\subsubsection{Topological energies for the real solutions of the Hermitian
counterpart}

At first we derive the Hamiltonian corresponding to $\mathcal{L}_{\text{b}}$
in the standard fashion by computing the conjugate canonical momenta%
\begin{equation}
\Pi ^{a}=\frac{\delta \mathcal{L}_{\text{b}}}{\delta \varphi _{0}^{a}}%
=G_{ac}\varphi _{0}^{c},~~\ \text{\ with \ }G_{ac}=\frac{\lambda ^{2}}{2}%
\left( \sin \zeta -\imath \epsilon \cos \zeta \right) ^{4}\sin ^{2}\Theta 
\mathcal{Q}_{a}^{i}\mathcal{Q}_{c}^{i},
\end{equation}%
so that%
\begin{equation}
\mathcal{H}_{\text{b}}=\frac{1}{2}\Pi ^{a}G_{ac}^{-1}\Pi ^{c}-\mathcal{L}_{%
\text{b}},~~\ \text{\ with \ }G_{ac}^{-1}=\frac{2\varphi _{i}^{a}\varphi
_{i}^{c}}{J^{2}\lambda ^{2}\left( \sin \zeta -\imath \epsilon \cos \zeta
\right) ^{4}\sin ^{2}\Theta },
\end{equation}%
where $J:=$ $\frac{1}{2}\varepsilon _{abc}\varepsilon ^{ijk}\varphi
_{i}^{a}\varphi _{j}^{b}\varphi _{k}^{c}$.

While overall our considerations are mainly classical, we now briefly appeal
to the quantum field theoretic version of the model, by assuming the
standard canonical equal time commutation relation $\left[ \varphi
^{a}(r,t),\Pi ^{b}(r^{\prime },t)\right] =i\delta ^{ab}\delta (r-r^{\prime
}) $ between the fields $\varphi ^{a}(r,t)$ and their conjugate momentum
operators $\Pi ^{a}(r,t)$. We then use a slightly modified version of the
Dyson operator as employed in \cite{Bender:2005hf,fring2020BPS}%
\begin{equation}
\eta =\exp \left[ -\func{arctanh}\epsilon \sum\nolimits_{a}\int dx\Pi
^{a}(r,t)\right] ,
\end{equation}%
to map the non-Hermitian Hamiltonian functional $\mathcal{H}_{\text{b}}$ to
a Hermitian counterpart $\mathfrak{h}_{\text{b}}$ by means of the adjoint
action of $\eta $ 
\begin{equation}
\mathfrak{h}_{\text{b}}=\eta \mathcal{H}_{\text{b}}\eta ^{-1}=\frac{1}{2}\Pi
^{a}G_{ac}^{-1}\Pi ^{c}+\frac{\tilde{\lambda}^{2}}{4}\sin ^{4}\zeta \sin
^{2}\Theta \mathcal{B}_{\mu }\mathcal{B}^{\mu }+\tilde{\mu}^{2}(1-\cos \zeta
).
\end{equation}%
We notice that $\mathfrak{h}_{\text{b}}$ is in fact the standard BPS Skyrme
model with reversing the previous re-scaling of the coupling constants as $%
\lambda \rightarrow \tilde{\lambda}=\lambda (1-\epsilon ^{2})$, $\mu
\rightarrow \tilde{\mu}=\mu (1-\epsilon ^{2})^{1/4}$.

In this case the static BPS solution that saturates the Bogomolny bound is
known to be computable exactly \cite{adam2010skyrme} when using spherical
space-time coordinates $(x,y,z)\rightarrow (r,\theta ,\phi )$ with $r\in
\lbrack 0,\infty )$, $\theta \in \lbrack 0,\pi )$, $\phi \in \lbrack 0,2\pi
) $ and the identifications $\Theta =\theta $, $\Phi =n\phi $ with $n\in 
\mathbb{Z}$ together with the assumption that $\zeta $ is a function of $r$
only. In this case one obtains a well-defined real compacton solution, see
e.g. \cite{comp4} for what that entails in general, 
\begin{equation}
\zeta _{r}(r)=\left\{ 
\begin{array}{ll}
2\arccos \left( \frac{1}{\sqrt{2}}\left\vert \frac{\tilde{\mu}}{n\tilde{%
\lambda}}\right\vert ^{1/3}r\right) ~~\ \  & \text{for }r\in \left[ 0,r_{c}=%
\sqrt{2}\left\vert \frac{n\tilde{\lambda}}{\tilde{\mu}}\right\vert ^{1/3}%
\right] \\ 
0 & \text{otherwise}%
\end{array}%
\right. ,
\end{equation}%
with real energy%
\begin{equation}
E=8\pi \tilde{\mu}^{2}\int\nolimits_{0}^{r_{c}}r^{2}V\left[ \zeta _{r}(r)%
\right] dr=\frac{64}{15}\sqrt{2}\left\vert n\right\vert \tilde{\mu}\tilde{%
\lambda}\pi (1-\epsilon ^{2})^{5/4}.  \label{ER}
\end{equation}%
Next we show that there are in fact more solutions in this case and how the
same energy results from a direct computation for the complex solution of
the non-Hermitian system (\ref{Lb}).

\subsubsection{Energies for the complex solutions of the non-Hermitian system%
}

We adopt here and below the approach proposed in \cite{adam2013some}, which
slightly reformulates the BPS theory and exploits the self-duality and
anti-self-duality between certain fields. For this purpose we first note
that the Hamiltonian density for static solutions may be expressed as 
\begin{equation}
\mathcal{H}_{\text{b}}=A^{2}+\tilde{A}^{2},
\end{equation}%
with%
\begin{equation}
A:=\frac{\lambda }{2}\left( \sin \zeta -\imath \epsilon \cos \zeta \right)
^{2}\sin \Theta \mathcal{B}_{0},~~\ \ \ \ \tilde{A}=\mu V=\mu \left( \sqrt{%
1-\epsilon ^{2}}-\cos \zeta -\imath \epsilon \sin \zeta \right) ^{1/2}.
\label{AD}
\end{equation}%
The self-duality and anti-self-duality between the fields $A$ and $\tilde{A}$
\begin{equation}
A=\pm \tilde{A},  \label{AAA}
\end{equation}%
is then interpreted as being identical to the BPS equations \cite%
{bogomol1976stability,prasad1975exact}. The energy functional for the
solutions of (\ref{AAA}) therefore acquires the form%
\begin{equation}
E_{\text{b}}=\int d^{3}x\left( A^{2}+\tilde{A}^{2}\right) =\pm 2\int d^{3}x~A%
\tilde{A}.  \label{Eb}
\end{equation}%
Note that since the fields $A$ and $\tilde{A}$ are now complex, the energy
of the static BPS solutions $E_{\text{b}}$ may no longer saturate the
Bogomolny bound.

Explicitly the BPS equations (\ref{AAA}) may be written as 
\begin{equation}
\frac{\lambda }{2}\frac{\left( \sin \zeta -\imath \epsilon \cos \zeta
\right) ^{2}}{\mu \sqrt{V}}\sin \Theta \varepsilon ^{ijk}\partial _{i}\zeta
\partial _{j}\Theta \partial _{k}\Phi =\pm 1.  \label{t1}
\end{equation}%
Since $\varepsilon _{ijk}\zeta _{i}\Theta _{j}\Phi _{k}$ is simply the
Jacobian for the variable transformation $(x,y,z)\rightarrow (\Theta ,\Phi
\,,\zeta )$ the multiplication of\ (\ref{t1}) by the volume element $d^{3}x$
leads to%
\begin{equation}
\frac{\lambda }{2}\frac{\left( \sin \zeta -\imath \epsilon \cos \zeta
\right) ^{2}}{\mu \sqrt{V}}\sin \Theta d\zeta d\Theta d\Phi =\pm r^{2}\sin
\theta drd\theta d\phi ,  \label{t2}
\end{equation}%
where we used spherical coordinates on the right hand side. With the same
identifications between $(r,\theta ,\phi )$ and $(\zeta ,\Theta ,\Phi )$ as
chosen in the previous section and together with the aforementioned
trigonometric identities the relation (\ref{t2}) converts into%
\begin{equation}
\frac{n\tilde{\lambda}}{2r^{2}}\sin ^{2}\left( \zeta -\imath \func{arctanh}%
\epsilon \right) \frac{d\zeta }{dr}=\pm \tilde{\mu}\sqrt{1-\cos \left( \zeta
-\imath \func{arctanh}\epsilon \right) }.  \label{bog}
\end{equation}%
These equations is easily integrated out by separating variables.
Corresponding to the different branches we obtain different types of
solutions 
\begin{equation}
\zeta _{i,m}^{\pm }(r)=\tilde{\zeta}_{i,m}^{\pm }(r)+\imath \func{arctanh}%
\epsilon =2\arccos \left[ \omega ^{i}\frac{(n\tilde{\lambda}c\mp \tilde{\mu}%
r^{3})^{1/3}}{\sqrt{2}n^{1/3}\tilde{\lambda}^{1/3}}\right] +2\pi m+\imath 
\func{arctanh}\epsilon ,~  \label{z1}
\end{equation}%
for $i=0,1,2$, $m\in \mathbb{Z}$ and $\omega =e^{2\pi \imath /3}$ denoting
the third root of unity. We analytically continue here the $\arccos $%
-function to the entire complex plane by the well-known formula $\arccos
z=-\imath \ln \left( z\pm \sqrt{z^{2}-1}\right) $. Note that for the
Hermitian case, i.e. $\epsilon =0$, all these solutions also arise, but in
that case one simply discards the complex solutions or the parts of the
solutions that become complex after a certain value of $r$, by requiring
solutions to be real. In order to identify possible compacton solutions in
the real part we need to specify the critical values $r_{0}$ for which the
solution vanish, $\tilde{\zeta}_{i}^{\pm }(r_{0})=0$, and also those values $%
r_{\pi }$ for which $\tilde{\zeta}_{i}^{\pm }(r_{\pi })=\pi $. We obtain 
\begin{equation}
r_{0,i}^{\pm }:=\omega ^{i}\left[ \frac{\pm n\tilde{\lambda}\left(
c-2^{3/2}\right) }{\tilde{\mu}}\right] ^{1/3}~,~\ \ \text{and}~~\ \ \ r_{\pi
,i}^{\pm }:=\omega ^{i}\left( \frac{\pm n\tilde{\lambda}c}{\tilde{\mu}}%
\right) ^{1/3}.  \label{z4}
\end{equation}%
\ \ These values are irrelevant when complex, whereas when real they may
produce different types of scenarios depending on their ordering and signs
of the constants. In figure \ref{FigStandardBPS} we depict some interesting
possibilities.

\FIGURE{ \epsfig{file=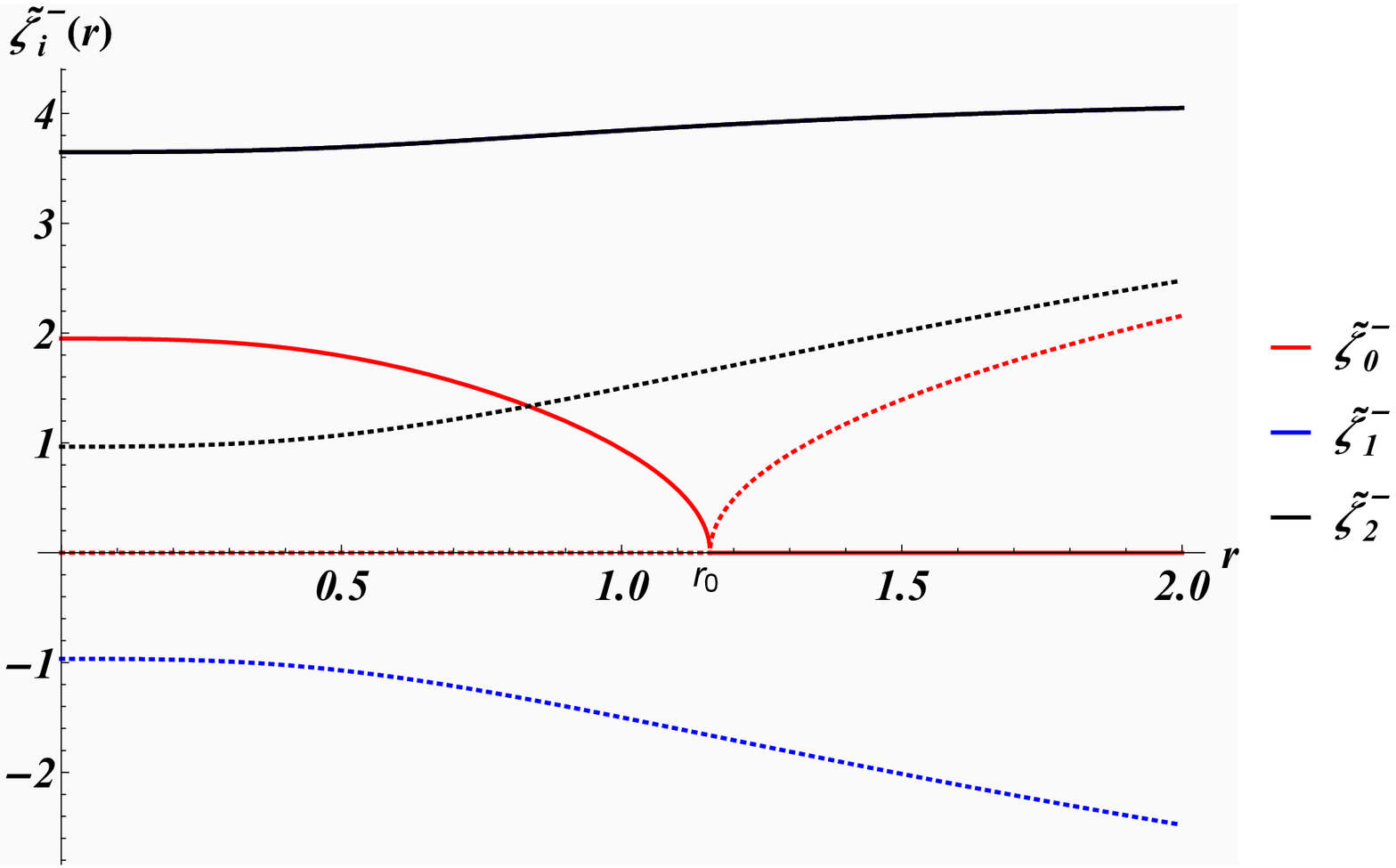, width=7.6cm} \epsfig{file=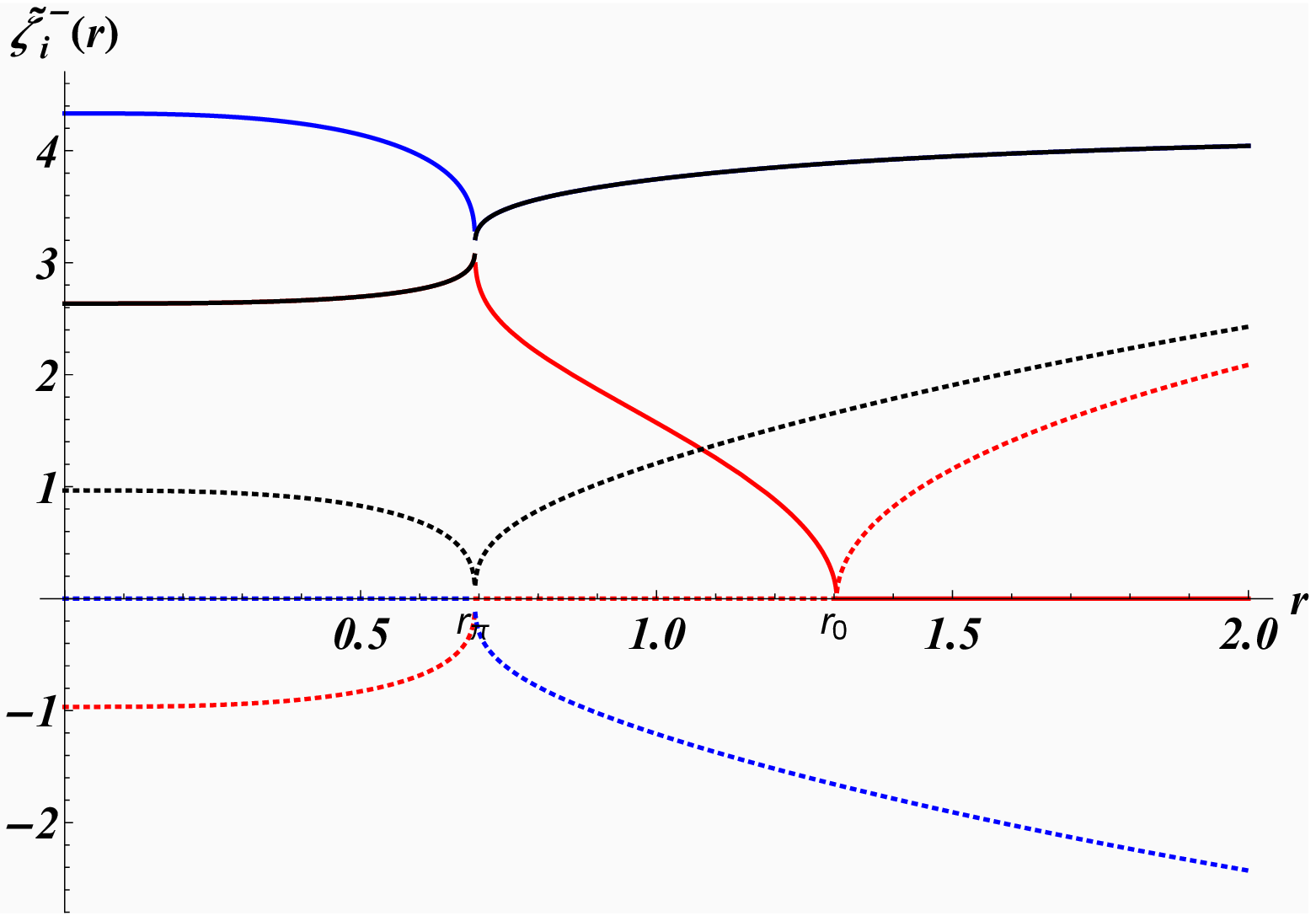,width=6.8cm}
\epsfig{file=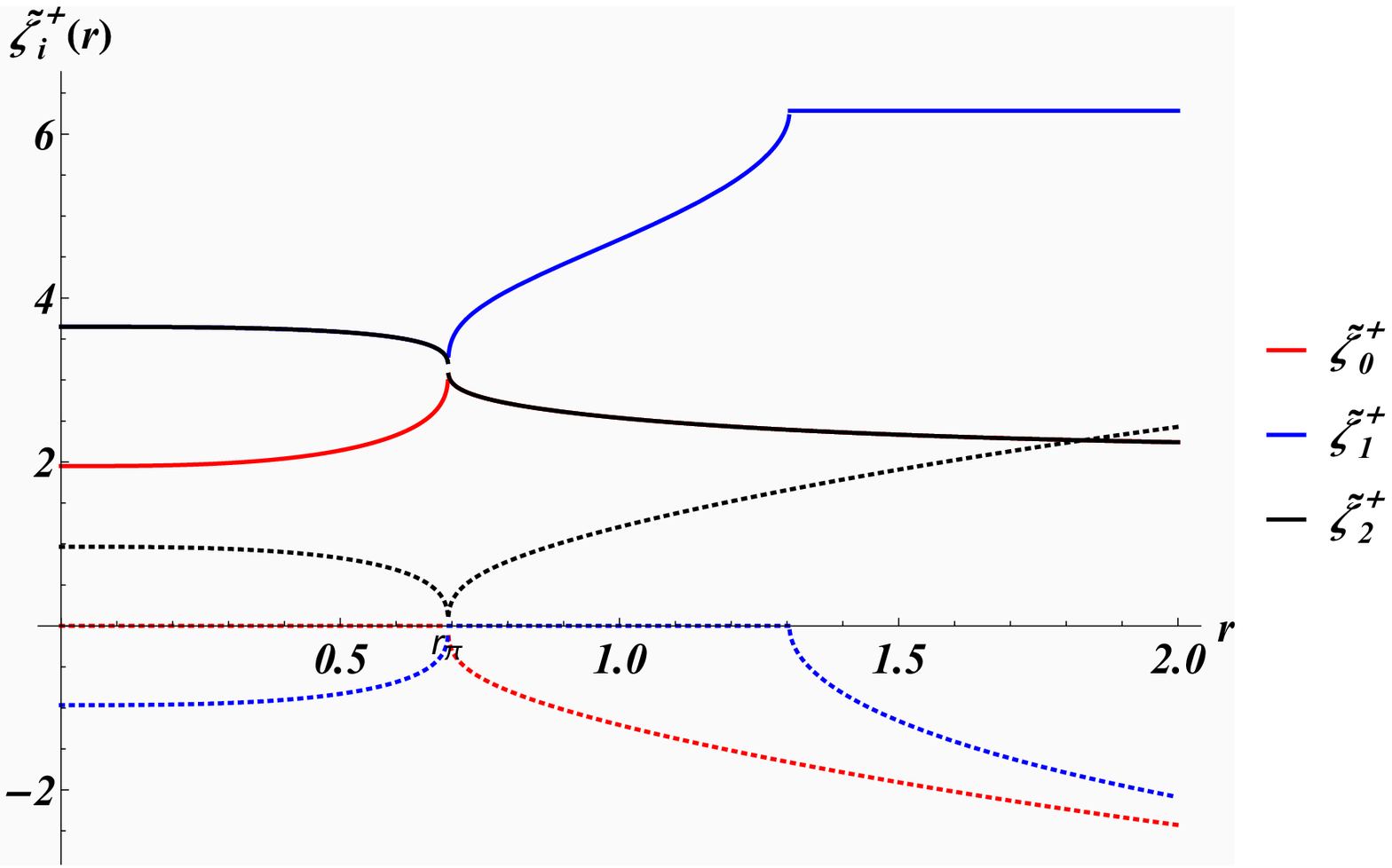, width=7.6cm} \epsfig{file=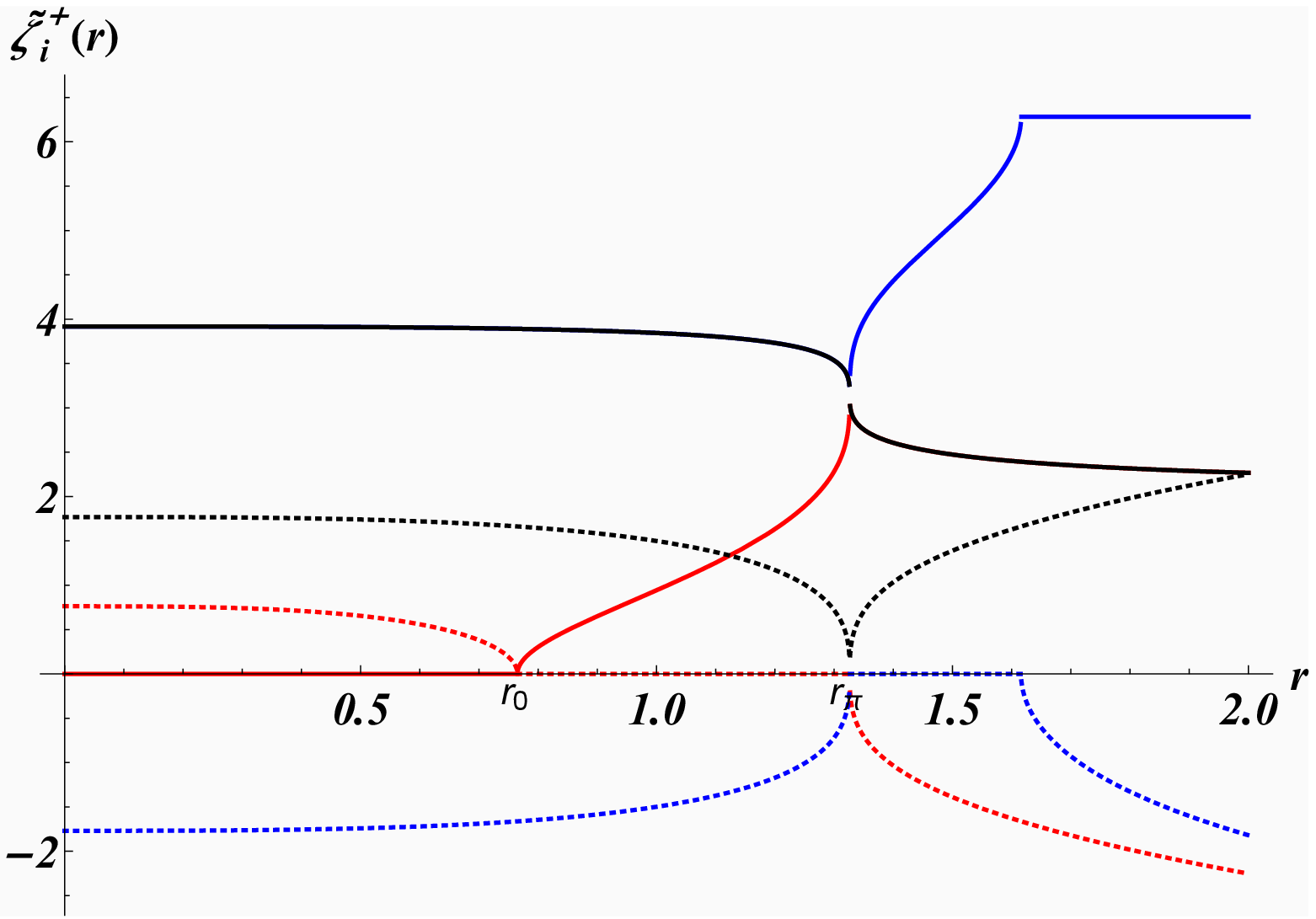,width=6.8cm}
\caption{The $\tilde{\zeta}$-part of the
solutions of the BPS equation (\ref{bog}) for different scenarios with $n=\tilde{\lambda}=1$, $\mu =3/2$ and different choices for $c$. The different relative orderings are: panel 
(a) $ r_{\pi } \leq 0\leq  r_{0} $ with $c=1/2$, panel (b) $0\leq 
r_{\pi} \leq $ $r_{0 }$ with $c=-1/2$, panel (c) $r_{0}\leq $ $0\leq r_{\pi }$ with 
$c=1/2$ and panel (d) $0\leq r_{0 }\leq r_{\pi}
$ with $c=7/2$. Real parts correspond to solid lines and imaginary parts to dotted ones.}
       \label{FigStandardBPS}}

It is clear from figure \ref{FigStandardBPS} that we may construct compacton
type solutions in various ways. Obvious choices are 
\begin{equation}
\tilde{\zeta}_{\text{BPS}}(r):=\left\{ 
\begin{array}{ll}
\tilde{\zeta}_{0,0}^{-}~~\  & \text{for \ \ }0\leq r\leq r_{0}^{-} \\ 
0 & \text{for \ \ }r_{0}^{-}<r%
\end{array}%
\right. ,~~~\ \ \text{\ }\tilde{\zeta}_{\text{St}}(r):=\left\{ 
\begin{array}{ll}
\tilde{\zeta}_{1,0}^{-}~~\  & \text{for \ \ }0\leq r\leq r_{\pi }^{-} \\ 
\tilde{\zeta}_{0,0}^{-} & \text{for \ \ }r_{\pi }^{-}\leq r\leq r_{0}^{-} \\ 
0 & \text{for \ \ }r_{0}^{-}<r%
\end{array}%
\right. .  \label{zeta2}
\end{equation}

Noting that $r_{\pi ,i}^{+}(c)=r_{\pi ,i}^{-}(-c)$, we may also glue
together solution that are self-dual with those that are anti-self-dual as 
\begin{equation}
\tilde{\zeta}_{\text{Cusp}}(r):=\left\{ 
\begin{array}{ll}
\tilde{\zeta}_{0,0}^{+}~~\  & \text{for \ \ }0\leq r\leq r_{\pi }^{+}=r_{\pi
}^{-} \\ 
\tilde{\zeta}_{0,0}^{-} & \text{for \ \ }r_{\pi }^{-}\leq r\leq r_{0}^{-} \\ 
0 & \text{for \ \ }r_{0}^{-}<r%
\end{array}%
\right. ,~~~\ \ \tilde{\zeta}_{\text{Shell}}(r):=\left\{ 
\begin{array}{ll}
0 & \text{for \ \ }r<r_{0}^{+} \\ 
\tilde{\zeta}_{0,0}^{+}~~\  & \text{for \ \ }r_{0}^{+}\leq r\leq r_{\pi
}^{+}=r_{\pi }^{-} \\ 
\tilde{\zeta}_{0,0}^{-} & \text{for \ \ }r_{\pi }^{-}\leq r\leq r_{0}^{-} \\ 
0 & \text{for \ \ }r_{0}^{-}<r%
\end{array}%
\right. .  \label{zeta1}
\end{equation}%
A purely imaginary compacton solution is obtained as $\tilde{\zeta}_{\text{%
iBPS}}(r):=\tilde{\zeta}_{0,0}^{+}$ for $r<r_{0}^{+}$ and $0$ otherwise. We
have dropped here the second subscript on $r_{0,i}^{\pm }$ and $r_{\pi
,i}^{\pm }$ as the branch that produces a real values depends on the values
of $\tilde{\lambda}$, $\tilde{\mu}$ and $c$. It appears that in this way one
is combining solutions from different equations. However, noting that the
equation of motion resulting from (\ref{Lb}) is simply the square of the BPS
equations (\ref{bog}), see e.g. \cite{adam2010skyrme} for a derivation when $%
\epsilon =0$, we adopt here the view that the latter is more fundamental.
Hence, we are combining solutions for one single equation with different
choices of integration constants in different domains. Whilst the first
order derivative are discontinuous at the `gluing points' $r_{0}^{\pm }$ \
and $r_{\pi }^{\pm }$ in the solutions in (\ref{zeta2}) and (\ref{zeta1}),
we may argue here in a similar way as in\ \cite{adam2010skyrme} to establish
that the solutions are in fact well defined solutions. The derivative $d%
\tilde{\zeta}/dr$ always occurs multiplied with a $\sin ^{2}\tilde{\zeta}$
in the BPS equations, so that the left and right limits of this combination
is always finite at the gluing points, but might differ by a sign. Since
this sign is irrelevant in the equations of motion the solutions are well
defined and lead to meaningful values for the energy density and the Baryon
number density. We depict the configurations (\ref{zeta1}) - (\ref{zeta2})
in figure \ref{combinedsol}.

\FIGURE{ \epsfig{file=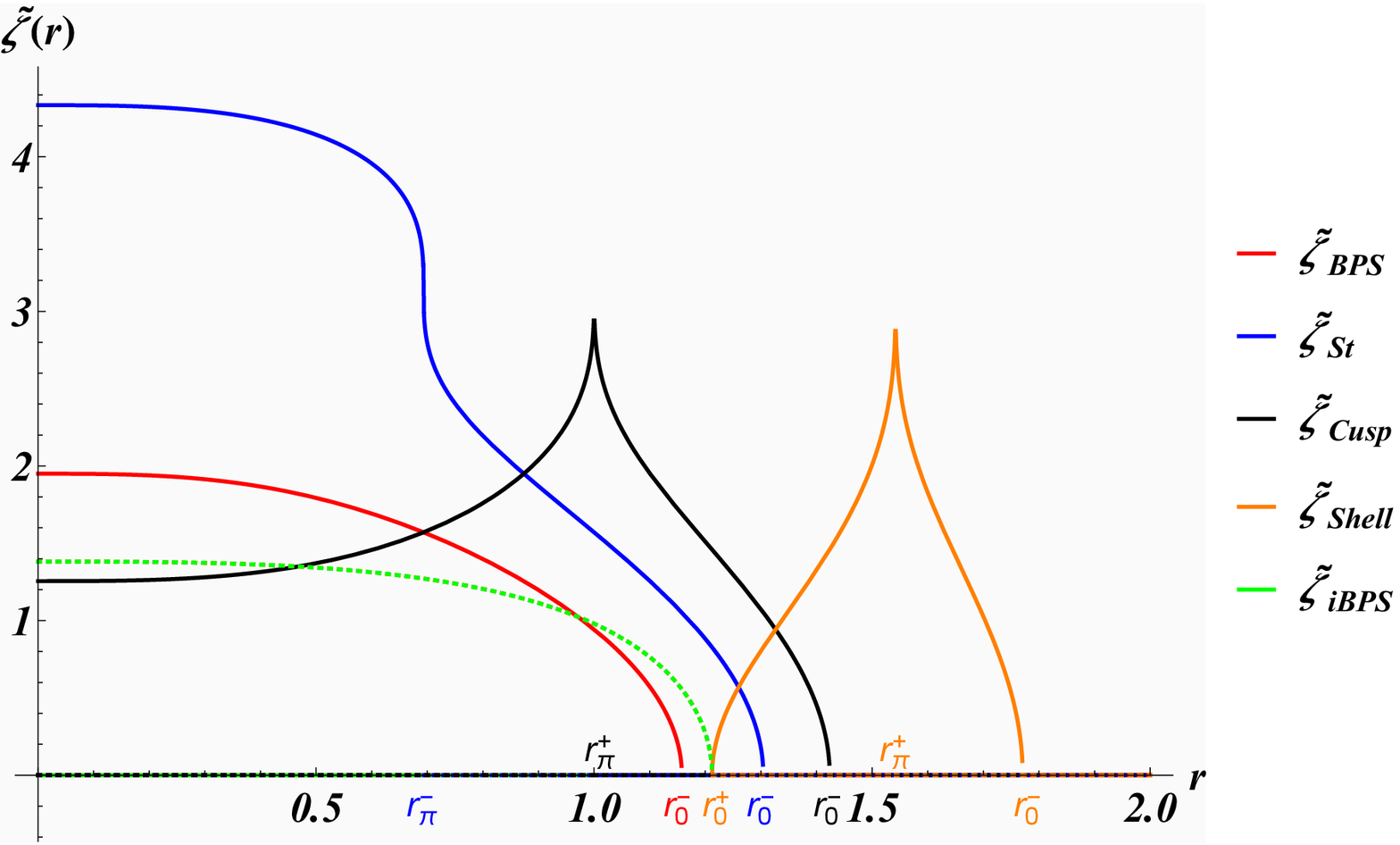, width=10.6cm} 
\caption{The BPS solution $\tilde{\zeta}_{\text{BPS}}$ with $n=1$, $c=1/2$, $\tilde{\mu}=3/2$, $\tilde{\lambda}=1$, the step solution $\tilde{\zeta}_{\text{St}}$
with $n=1$, $-c=1/2$, $\tilde{\mu}=3/2$, $\tilde{\lambda}=1$, the cusp
solution $\tilde{\zeta}_{\text{Cusp}}$ with $n=1$, $c=3/2$, $\tilde{\mu}=3/2$, $\tilde{\lambda}=1$, the shell solution $\tilde{\zeta}_{\text{Shell}}$
with $n=1$, $c=11/2$, $\tilde{\mu}=3/2$, $\tilde{\lambda}=1$ and the purely
imaginary solution $\tilde{\zeta}_{\text{iBPS}}$ with $n=1$, $c=11/2$, $\tilde{\mu}=3/2$, $\tilde{\lambda}=1$.}
       \label{combinedsol}}

In figure \ref{Differenttypes} we present the Skyrmion solutions of
compacton type (\ref{zeta1}) - (\ref{zeta2}) as slices in form of level
curves. We may compare with figure \ref{combinedsol}. In panel (a) we have a
standard real (fractional) Skyrmion $\tilde{\zeta}_{\text{BPS}}$ starting at
a finite value at $r=0$ and then decaying to zero at some critical value $%
r_{0}^{-}$. In panel (b) we depict the solution $\tilde{\zeta}_{\text{St}}$
taking on the form of a step like function with an inflection point at $%
r_{\pi }^{-}$. The solution $\tilde{\zeta}_{\text{Cusp}}$ shown in panel (c)
has a discontinuous first order derivative at $r=r_{\pi }^{+}=r_{\pi }^{-}$,
which is usually referred to as peakons in the context of 1+1 dimensional
integrable systems. The most interesting structure$\ \tilde{\zeta}_{\text{%
Shell}}$ is seen in panel (d), which corresponds to a real shell with a
peakon structure. We may even change this solution in the region $%
r<r_{0}^{+},$ by defining it as $\ \tilde{\zeta}_{\text{Core}}(r)=\tilde{%
\zeta}_{0,0}^{+}$ for $r<r_{0}^{+}$, hence adding a purely imaginary core to
it. It turns out that this is consistent as the core has also real energies
despite the fact that it is complex.

\FIGURE{ \epsfig{file=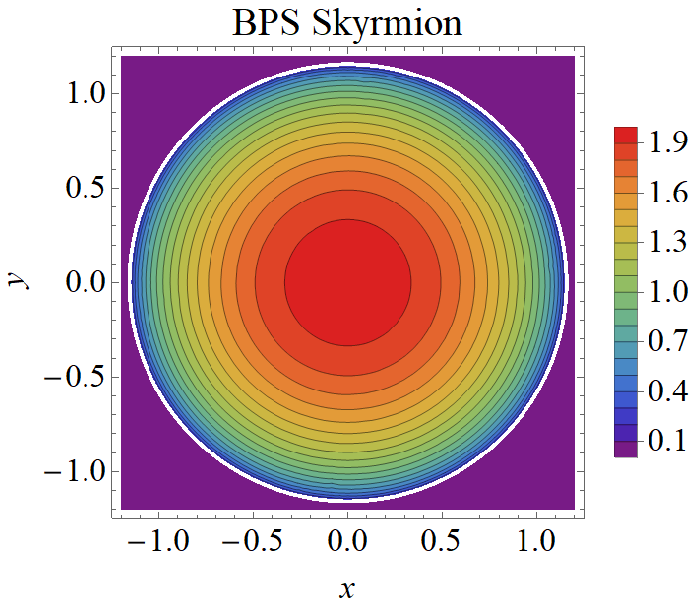, width=3.6cm} \epsfig{file=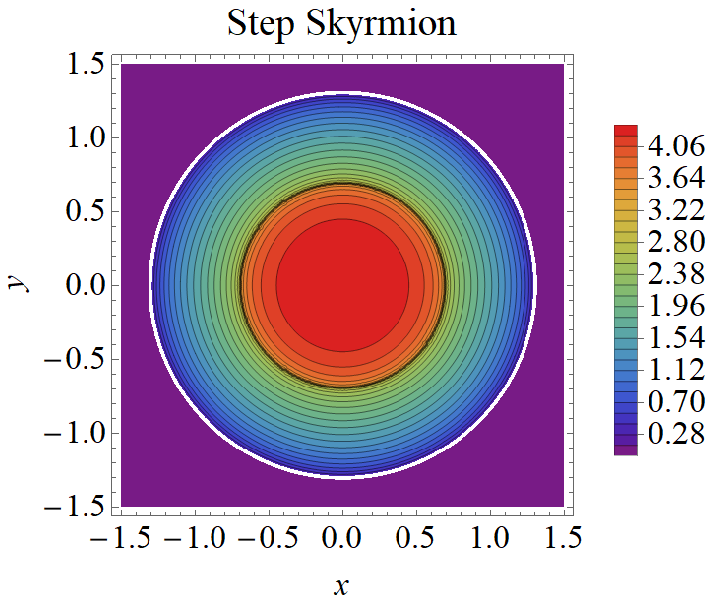, width=3.6cm}  \epsfig{file=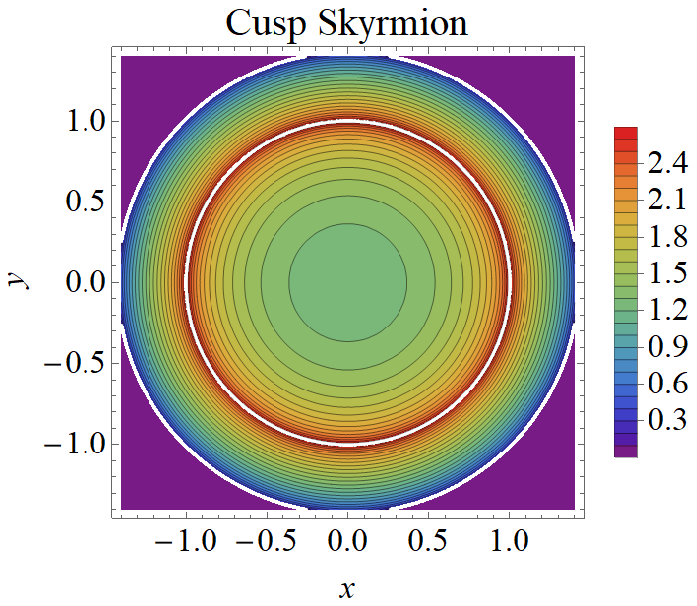, width=3.6cm} \epsfig{file=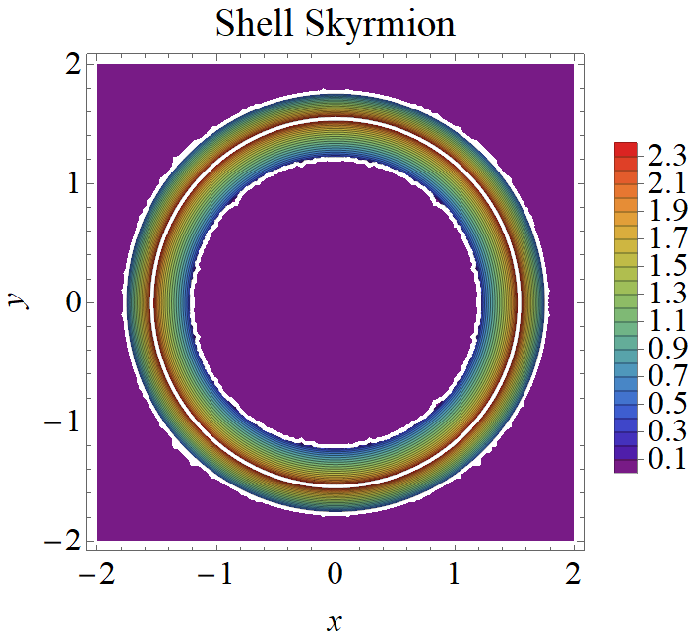, width=3.6cm}
\caption{Different types of solutions to the equations of motion as defined in (\ref{zeta1}) - (\ref{zeta2}) with parameters $n=\tilde{\lambda}=1$, $\tilde{\mu}=3/2$ and $c=1/2$ in panel (a), $c=-1/2$ in panel
(b), $c=3/2$ in panel (c), $c=11/2$ in panel (d).}
       \label{Differenttypes}}

Next we demonstrate that all types of solutions depicted in figures \ref%
{combinedsol} and \ref{Differenttypes} possess real energies. We compute
these energies on some domain $r\in \lbrack \tilde{r}_{c},r_{c}]$ by using
the general expression (\ref{Eb})%
\begin{eqnarray}
E &=&\pm \tilde{\lambda}\tilde{\mu}\int d^{3}x\left( \sin \zeta -\imath
\epsilon \cos \zeta \right) ^{2}\sin \Theta \mathcal{B}_{0}\left( \sqrt{%
1-\epsilon ^{2}}-\cos \zeta -\imath \epsilon \sin \zeta \right) ^{1/2}
\label{En} \\
&=&\pm 4\pi \tilde{\lambda}\tilde{\mu}n\int\nolimits_{\tilde{r}%
_{c}}^{r_{c}}dr\sin ^{2}\left( \zeta (r)-\imath \func{arctanh}\epsilon
\right) \sqrt{1-\cos \left( \zeta (r)-\imath \func{arctanh}\epsilon \right) }%
\frac{d\zeta }{dr}~~~~~~~~ \\
&=&8\pi \tilde{\mu}^{2}\int\nolimits_{\tilde{r}_{c}}^{r_{c}}drr^{2}V\left[
\zeta (r)\right] .  \label{En3}
\end{eqnarray}%
In the last step we used once more equation (\ref{bog}). For the solutions $%
\tilde{\zeta}_{\text{BPS}}$, $\tilde{\zeta}_{\text{St}}$~and $\tilde{\zeta}_{%
\text{Cusp}}$ we calculate%
\begin{equation}
E_{\text{BPS/St,Cusp}}=\frac{8}{15}n\tilde{\mu}\tilde{\lambda}\pi \left( 8%
\sqrt{2}\mp 10c\pm 3c^{5/3}\right) ,  \label{Ebb}
\end{equation}%
for $c\geq 0$ on the domains as indicated in figure \ref{FigStandardBPS}.
The upper signs stand here for BPS and lower signs for the step and cusp
solutions, which have the same energies. As expected, the expressions (\ref%
{Ebb}) reduce to the energy of the standard real case (\ref{ER}) in the
limit $c\rightarrow 0$, since in that case the fractional BPS Skyrmions
become full BPS Skyrmions with $\zeta (r=0)=\pi $. For the shell solution $\ 
\tilde{\zeta}_{\text{Shell}}$ and the purely imaginary core solution $\tilde{%
\zeta}_{\text{iBPS}}$ we obtain the real energies%
\begin{equation}
E_{\text{Shell}}=\frac{128}{15}\sqrt{2}n\tilde{\mu}\tilde{\lambda}\pi ,~\ \ 
\text{and}~\ \ \text{ }E_{\text{iBPS}}=-E_{\text{BPS}},
\end{equation}%
respectively. The reality of the solutions is ensured by verifying that the
respective solutions satisfy all three conditions (\ref{cpt1})-(\ref{cpt3})
for a particular $\mathcal{CPT}^{\prime }$-symmetry. With condition (\ref%
{cpt1}) we identify here the symmetry to 
\begin{equation}
\mathcal{CPT}^{\prime }:\zeta (x_{\mu })\rightarrow \zeta ^{\ast }(-x_{\mu
})+2\imath \func{arctanh}\epsilon =\zeta (-x_{\mu }).  \label{cptd}
\end{equation}%
We are considering static solutions in which the angle dependence has
already been eliminated, so that our solutions only depend on $r$. Hence the
change in the arguments of the fields $x_{\mu }\rightarrow -x_{\mu }$ is
automatically satisfied. The $\mathcal{CPT}^{\prime }$-symmetry condition (%
\ref{cptd}) is then easily verified for our solutions $\zeta _{i,m}^{\pm
}(r) $ in (\ref{z1}): $\zeta _{i,m}^{\pm }(r)\rightarrow \left[ \zeta
_{i,m}^{\pm }(r)\right] ^{\ast }+2\imath \func{arctanh}\epsilon =\zeta
_{i,m}^{\pm }(r)$. Since the solutions are mapped to themselves, the
condition (iii) is automatically satisfied and energies for these solutions
must be real. Notice that the symmetry $\mathcal{CPT}^{\prime }$ differs
from the symmetry $\mathcal{CPT}$ we used for the construction of the model.

We conclude this section with a brief comment on the values for the Baryon
number, that in general is no longer integer valued. Taking the
normalization constant to be $N_{0}=24\pi ^{2}$ we obtain 
\begin{equation}
B=\int B_{0}d^{3}x=\frac{n}{\pi }\left[ \zeta ^{\pm }(r=0)-\frac{1}{2}\sin %
\left[ 2\zeta ^{\pm }(r=0)\right] \right] ,
\end{equation}%
which is no longer integer valued.

It is worth pointing out that we may reach similar conclusions as in the
boosted model discussed in this section for a model with complex rotated
fields. With a slight modification of the Dyson map used in \cite%
{fring2020goldstone}, having the effect on the fields is that they transform
as $\varphi ^{a}\rightarrow e^{-i\theta _{a}}\varphi ^{a}$ and $\Pi
^{a}\rightarrow e^{i\theta _{a}}\Pi ^{a}$, we may construct a new complex
models. The model obtained in this manner also possess complex BPS solutions
with real energies.

\section{Skyrme model with semi-kink and massless solutions}

While most Skyrmion solutions are of compacton type, there exist also
interesting variants of the model $\mathcal{L}_{0}+\mathcal{L}_{6}$ with
potentials that lead to solutions which are partly of kink type with real
energies. We consider here the potential%
\begin{equation}
V_{SK}(\zeta )=\sin ^{2}\zeta (1+\cos \zeta )^{2}.\ 
\end{equation}%
The corresponding BPS equations%
\begin{equation}
\tan \left( \frac{\zeta }{2}\right) \frac{d\zeta }{dr}=\pm \frac{2\mu }{%
n\lambda }r^{2},
\end{equation}%
are easily solved to%
\begin{equation}
\zeta _{s}^{\pm }(r)=2s~\arccos \left( e^{\mp \frac{\mu r^{3}}{3n\lambda }%
-c}\right) ,  \label{cksol}
\end{equation}%
with $s=\pm 1$ and $c$ denoting an integration constant. A similar solutions
to $s=1$ was found in \cite{bonen2010}. Evidently we have $\zeta _{s}^{\pm
}(r_{0}^{\pm })=0$ for $r_{0,i}^{\pm }=\omega ^{i}(\mp 3n\lambda c/\mu
)^{1/3}$ and asymptotically $\zeta _{s}^{\pm }$ acquires a finite value $%
\lim_{r\rightarrow \infty }\zeta _{s}^{\pm }(r)=s\pi $ for $\pm \mu
/n\lambda >0$. We depict some sample solutions in figure \ref{Figkink} panel
(a). For $r<r_{0}$ we notice the previously observed standard real or purely
imaginary compacton solutions, but for $r>r_{0}$ the solutions $\zeta _{\pm
}^{+}$ exhibit the interesting feature of being of compacton type at $%
r=r_{0} $ and of kink type when $r\rightarrow \infty $.

\FIGURE{ \epsfig{file=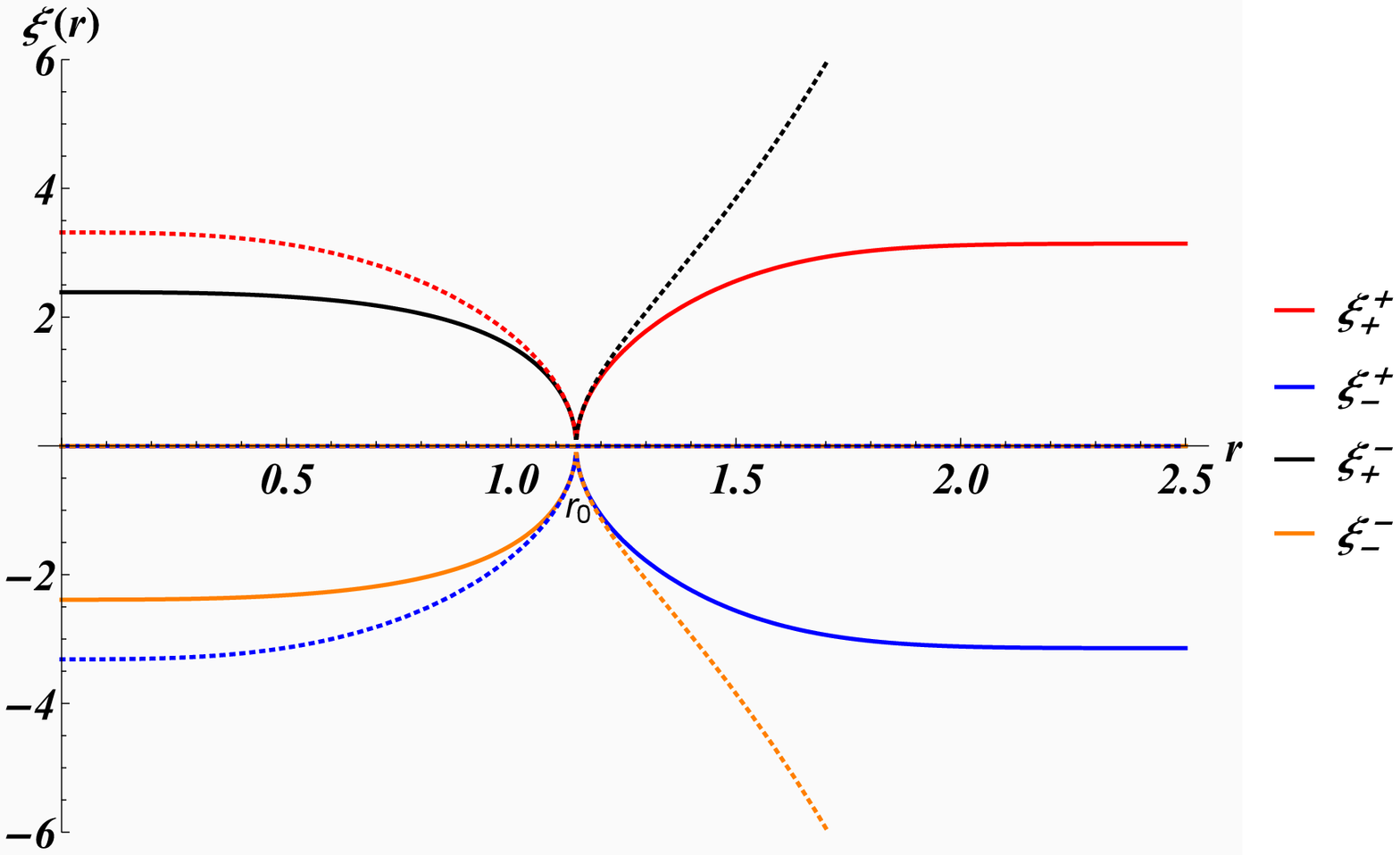, width=7.2cm} \epsfig{file=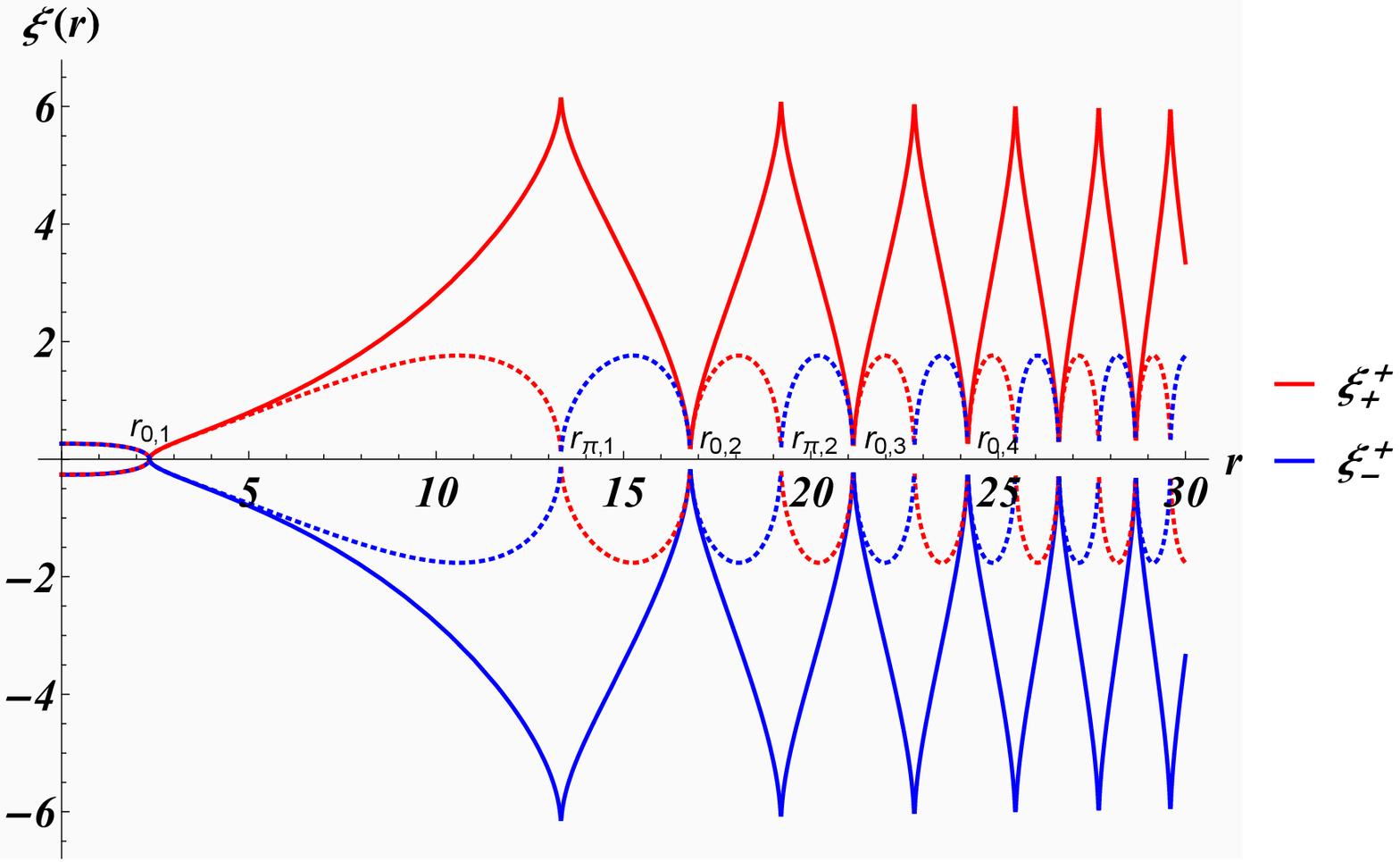, width=7.2cm}
\caption{Panel (a): Purely
imaginary and real compacton and semi-kink solutions (\ref{cksol}) resulting
from the potential $V_{SK}(\zeta )$ with parameter choices $n=\lambda =1$, $\mu =2$ and $c=\pm 1$ for $\gamma =\mp 1$. Panel (b): Complex solutions with zero energy resulting
from the potential $V_{m0}(\zeta )$ with parameter choices $n=\lambda =10$, $\mu =2$ and $c=\pm 25.15$ for $\gamma =\mp 1$. Real parts correspond to solid
lines and imaginary parts to dotted ones.}
       \label{Figkink}}

Crucially, it turns out that the energies of these solutions are all real
and finite. From the general expression (\ref{En3}) we compute 
\begin{eqnarray}
E_{\text{semi-kink}}\left( \zeta _{s}^{-}\right) &=&\frac{16n\lambda \mu \pi 
}{3}, \\
E_{\text{real compacton}}\left( \zeta _{s}^{-}\right) &=&-\left(
4e^{-6c}-3e^{-8c}-1\right) E_{\text{semi-kink}}, \\
E_{\text{purely imaginary compacton}}\left( \zeta _{s}^{+}\right) &=&\left(
4e^{6c}-3e^{8c}-1\right) E_{\text{semi-kink}}\text{,}
\end{eqnarray}%
for $c>0$ and $n\lambda \mu >0$.

Another interesting variant emerges when considering the potential $%
V_{m0}(\zeta )=-V_{SK}(\zeta )$. In this case the solutions become $\check{%
\zeta}_{s}^{\pm }(r)=2s~\arccos (e^{\mp \imath \frac{\mu r^{3}}{3n\lambda }%
-\imath c})$, which vanish for $r_{0,i}^{\pm }=\omega ^{i}(\mp 3n\lambda
(c+2\pi m)/\mu )^{1/3}$ with $m\in \mathbb{Z}$ and $\check{\zeta}_{s}^{\pm
}(r_{\pi ,i}^{\pm })=2\pi s$ for $r_{\pi ,i}^{\pm }=\omega ^{i}[\mp
3n\lambda (c+2\pi (m+1/2))/\mu ]^{1/3}$. A sample solution is depicted in
figure \ref{Figkink} panel (b). We observe a re-occurring complex periodic
shell solution that becomes squeezed for increasing $r$. Interestingly the
energies for these type of shell solutions is vanishing 
\begin{equation}
8\pi \mu ^{2}\int\nolimits_{r_{0}^{+}}^{r_{0}^{-}}drr^{2}V_{m0}\left[ \check{%
\zeta}_{s}^{\pm }(r)\right] =0.
\end{equation}%
This suggests that the shell solutions may be interpreted as massless
Skyrmions.

We observe from (\ref{cpt3}) that the energies of the solutions are ensured
to be real by the $\mathcal{CPT}_{\pm }$-symmetries: $\zeta (r)\rightarrow
\pm \zeta ^{\ast }(r)$. For the same reasons as in the previous subsection
there is no effect on the arguments of the fields. For the complex solution $%
\check{\zeta}_{s}^{\pm }$ this reads $\mathcal{CPT}_{\pm }$: $\zeta
_{s}^{\pm }(c)\rightarrow \left[ \zeta _{\pm s}^{\pm }(c)\right] ^{\ast
}=\zeta _{\pm s}^{\mp }(-c)$. Thus in this case this $\mathcal{CPT}_{\pm }$%
-symmetries map solutions to different solutions. However, invoking
condition (\ref{cpt3}) and noting that the energies for $\check{\zeta}%
_{s}^{\pm }(r)$ are the same for both BPS equations and independent of $s,c$%
, they must be real.

\section{Skyrme model with a Bender-Boettcher type potential}

We will now investigate further variants of the model $\mathcal{L}_{0}+%
\mathcal{L}_{6}$ by allowing for a wider range of potentials in $\mathcal{L}%
_{0}$, including the possibilities of functions of $\limfunc{Tr}\left[
U(\zeta )\right] $ and even $\limfunc{Tr}\left[ U(\imath \zeta )\right] $.
As a first example we consider the potential 
\begin{equation}
V_{BB}(\zeta )=\left( \imath \zeta \right) ^{\varepsilon }\sin ^{4}\zeta
,~~\ \varepsilon \in \mathbb{R}.\   \label{BB}
\end{equation}%
This potential closely resembles the classical prototype potential studied
in $\mathcal{PT}$-symmetric quantum mechanics \cite{Bender:1998ke},
remaining invariant under the $\mathcal{CPT}$-transformation: $\zeta
\rightarrow -\zeta $, $\imath \rightarrow -\imath $. Using the same
parameterization and reasonings as in the previous sections, the BPS
equations derived in analogy to (\ref{bog}) read 
\begin{equation}
\frac{n\lambda }{2\mu }\frac{\sin ^{2}\left( \zeta \right) }{\sqrt{V_{BB}}}%
d\zeta =\pm r^{2}dr~~~\ \Rightarrow ~~~\frac{d\zeta }{dr}=\pm \frac{2\mu }{%
n\lambda }\left( \imath \zeta \right) ^{\varepsilon /2}r^{2}~.  \label{int}
\end{equation}%
These equations are easily integrated, acquiring the following Gaussian form%
\begin{equation}
\zeta _{m}^{\pm }(r)=\left[ \left\vert \frac{n\lambda }{\mu (\varepsilon -2)}%
\right\vert \frac{1}{(c+r^{3}/3)}\right] ^{\frac{2}{\varepsilon -2}%
}e^{\imath \pi \left( \frac{3s}{2-\varepsilon }-\frac{1}{2}\right) }e^{2\pi
\imath \frac{2m}{\varepsilon -2}}\text{,~~~~\ }s=\pm 1,c\in \mathbb{R},m\in 
\mathbb{Z}\text{~}.  \label{sol}
\end{equation}%
In principle the integration constant $c~$could be complex, but we only
obtain real energies for $c\in \mathbb{R}$ so we ignore that possibility in
what follows. We have defined the constant $s:=\func{sign}[\pm n\lambda /\mu
(\varepsilon -2)]$ where as above $\func{sign}$ denotes the signum function.
The last factor accounts for all the branches of $\zeta $, as can either be
seen by inserting $1=e^{2\pi \imath m}$ into the square bracket or by noting
that $\zeta \rightarrow \zeta e^{2\pi \imath \frac{2m}{\varepsilon -2}}$ is
a symmetry of equation (\ref{int}). The BPS solutions $\zeta _{m}^{\pm }(r)$
exhibit two different types of qualitative behaviour. When $c\in \mathbb{R}%
^{-}$, $\varepsilon <2$ we obtain compacton solutions with finite values at $%
r=0$ and $\zeta _{m}^{\pm }[(3\left\vert c\right\vert )^{1/3}]=0$. For $c\in 
\mathbb{R}^{+}$, $\varepsilon >2$ the solutions are finite at $r=0$ and tend
to zero only for $r\rightarrow \infty $. We illustrate these types of
behaviour in figure \ref{Figzeta}.

\FIGURE{ \epsfig{file=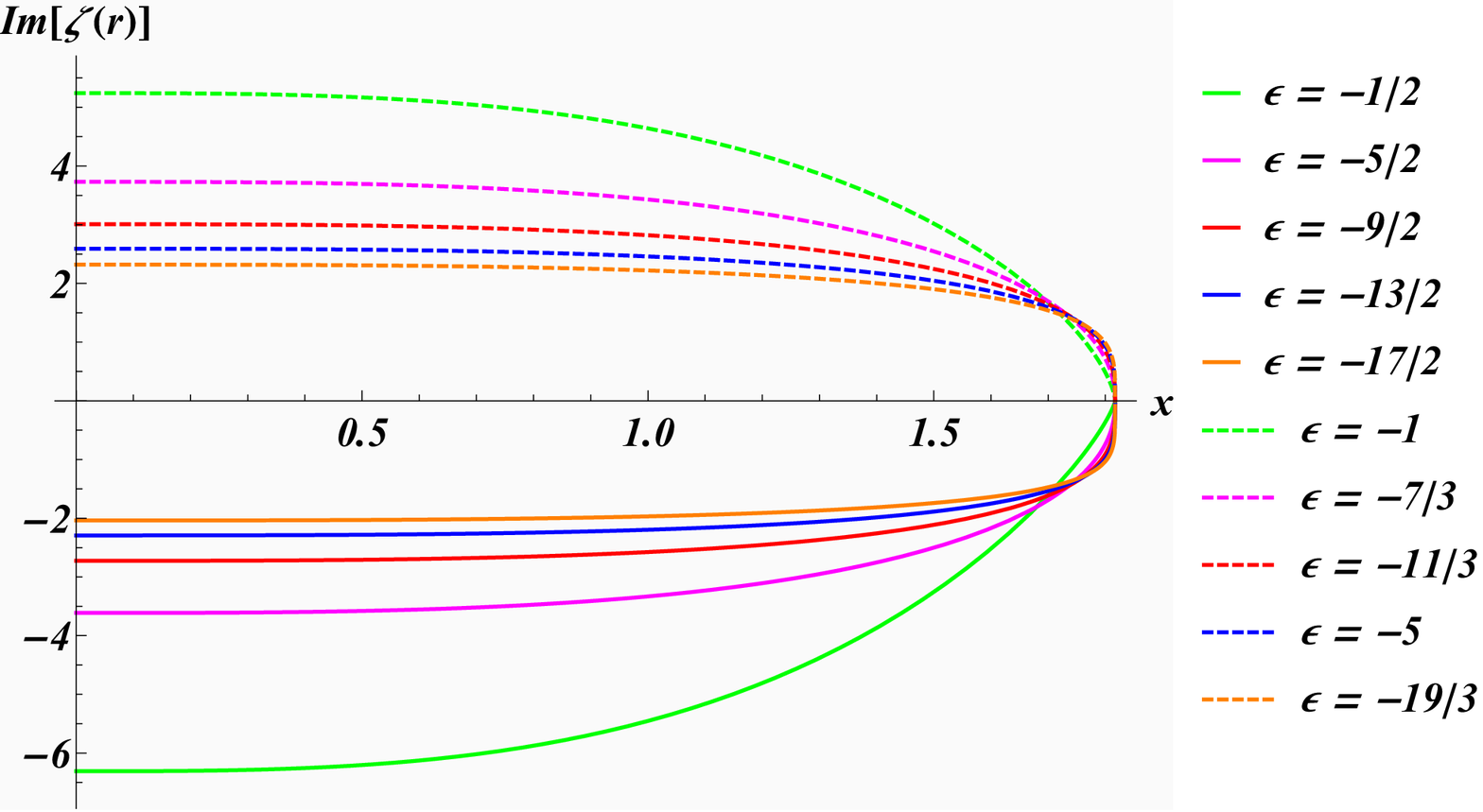, width=7.2cm} \epsfig{file=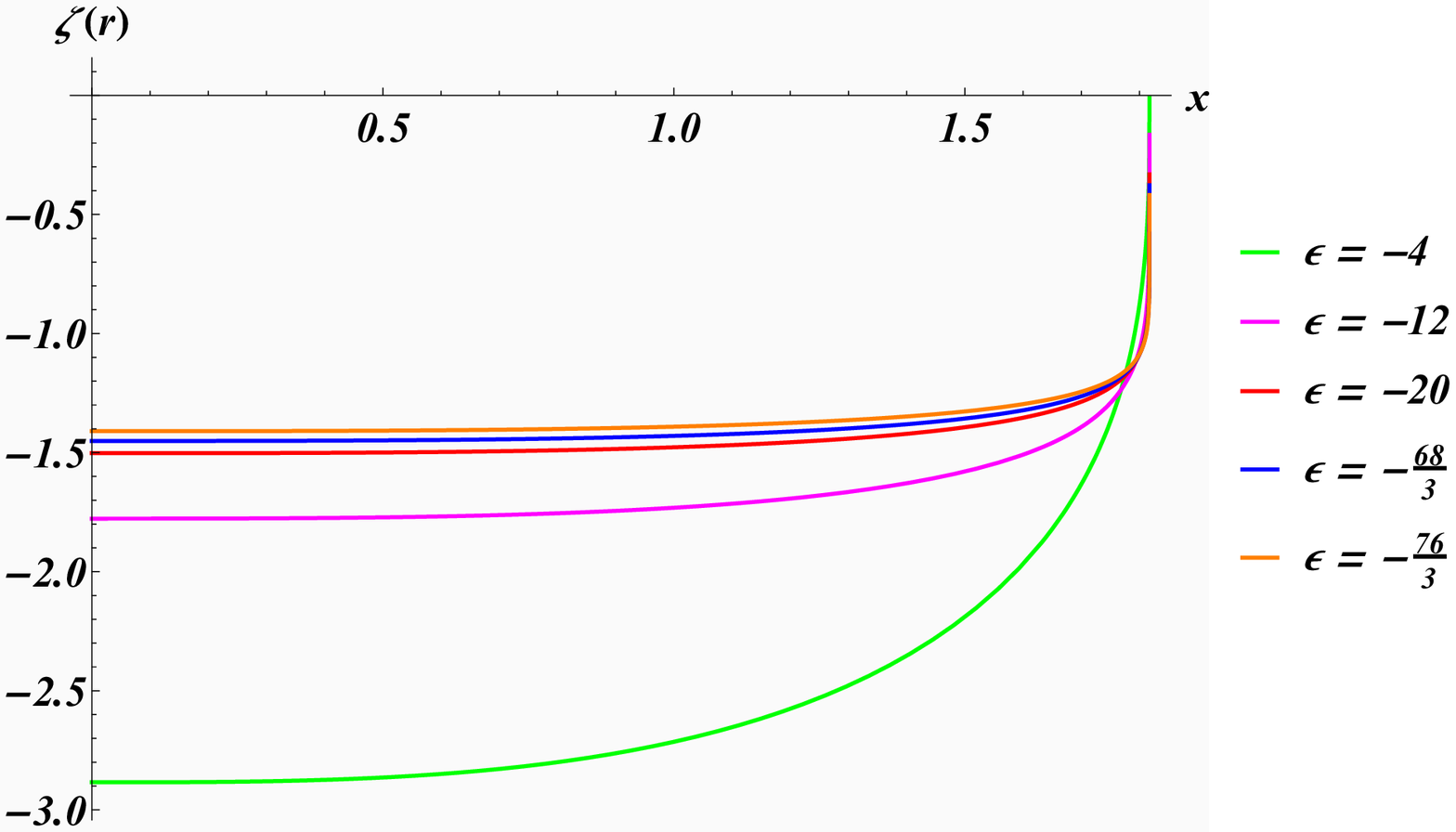,width=7.2cm}
\epsfig{file=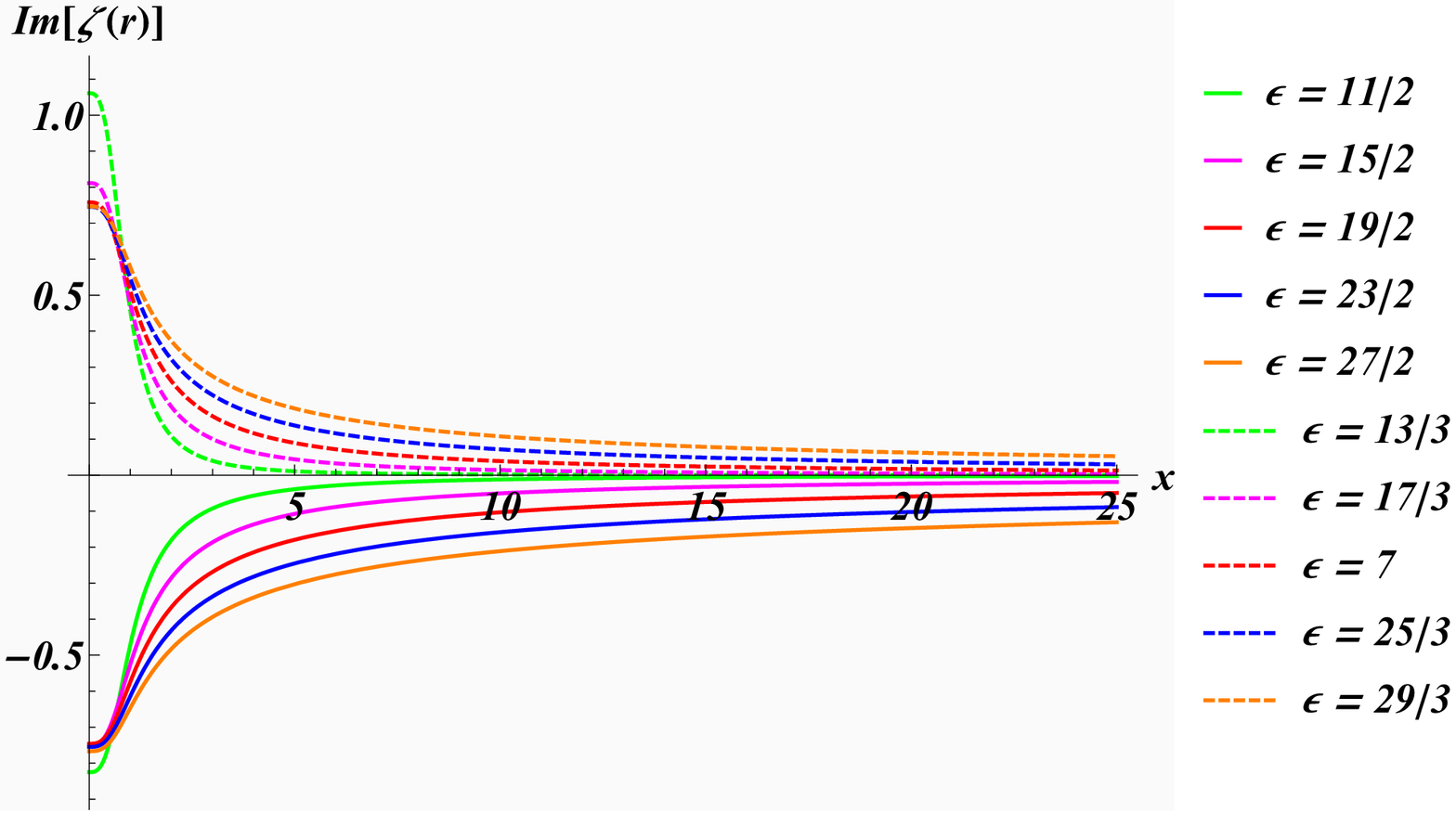, width=7.2cm} \epsfig{file=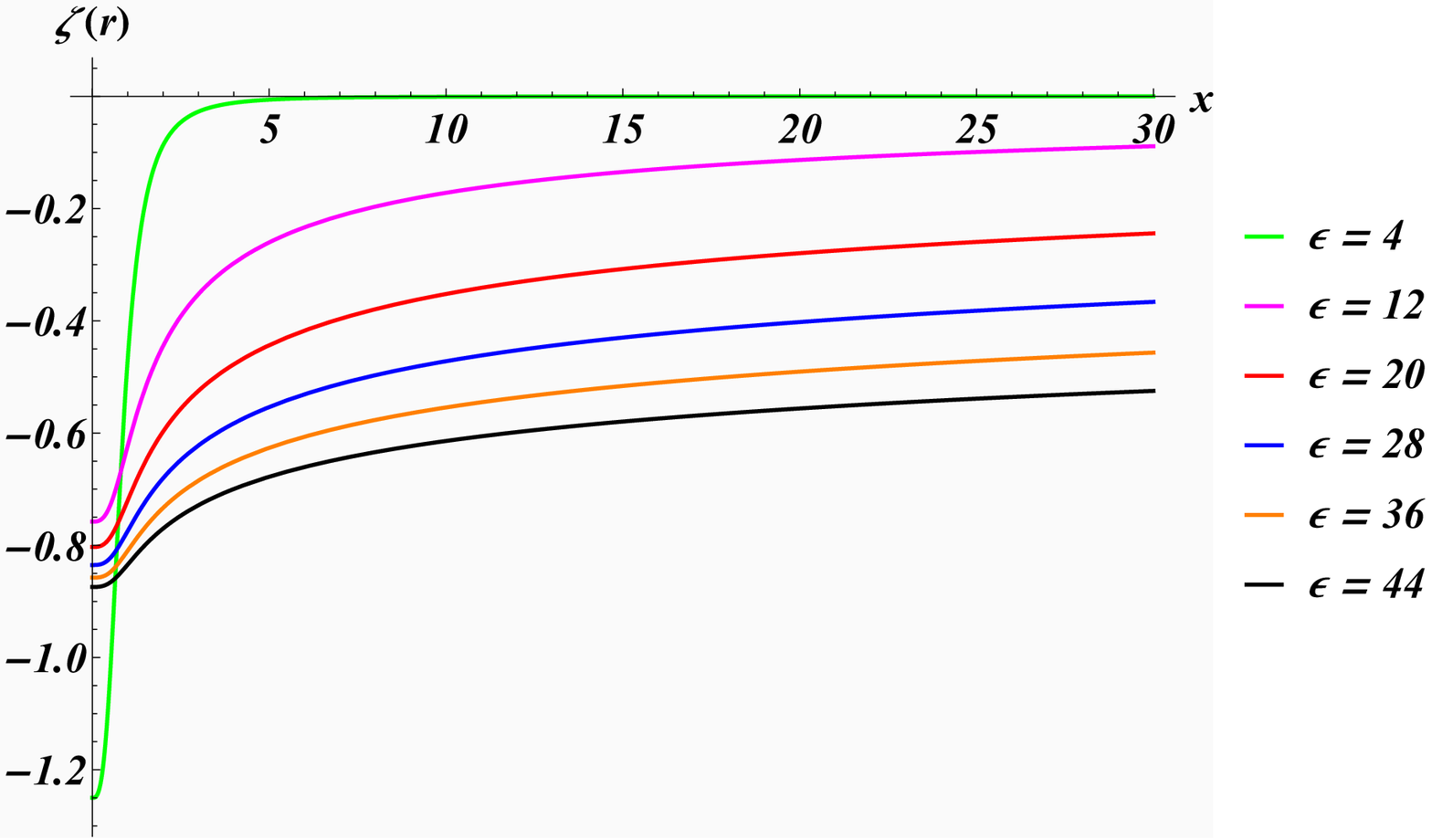,width=7.2cm}
\caption{BPS solutions $\zeta _{m}^{\pm }(r)$ for a Skyrme model with
a Bender-Boettcher type potential for the parameter choices $\lambda =1$, $\mu =2$, $n=1$. In panels (a), (b) we have taken $c=-2$ and  in panels (c), (d) we
have $c=0.2$. }
       \label{Figzeta}}

By the same reasoning as in the previous subsections the energies for these
solutions are computed to 
\begin{equation}
E_{\text{BB}}^{\pm }=8\pi \mu ^{2}\int\nolimits_{0}^{r_{c}}drr^{2}V\left[
\zeta _{m}^{\pm }(r)\right] ,
\end{equation}%
where $r_{c}=(3\left\vert c\right\vert )^{1/3}$ for the compacton solutions
and $r_{c}\rightarrow \infty $ for the unbounded ones. As is evident from (%
\ref{BB}) these energies can be real when $\zeta $ is either purely
imaginary or real. Together with (\ref{sol}) real energies are found when%
\begin{eqnarray}
&&\zeta \in -\imath \mathbb{R}^{+}\text{:\quad }\varepsilon =\frac{4m+4\ell
-3s}{2\ell }\text{,\quad }\ell ,m\in \mathbb{N}\text{,}  \label{c1} \\
\text{~} &&\zeta \in \imath \mathbb{R}^{+}\text{: \ \quad }\varepsilon =%
\frac{2+4m+4\ell -3s}{1+2\ell }\text{,\quad }m,\varepsilon \in \mathbb{N}%
\text{,}\ell \in \mathbb{N}_{0}\text{,}  \label{c2} \\
&&\zeta \in \mathbb{R}\text{: \ \quad }\varepsilon =\frac{2(1+4m+2\ell -3s)}{%
1+2\ell }\text{,\quad }\ell ,m\in \mathbb{N}_{0}\text{,}\ell ,\varepsilon
\in 4\mathbb{N}\text{.}  \label{c3}
\end{eqnarray}%
Examples for these solutions are depicted in figure \ref{Figzeta}. In panel
(b) of that figure we also displayed a two solution real solutions with $%
\varepsilon \notin 4\mathbb{N}$. Next we plot the corresponding energies for
these cases as functions of $\varepsilon $ in figure \ref{Fig1}.

\FIGURE{ \epsfig{file=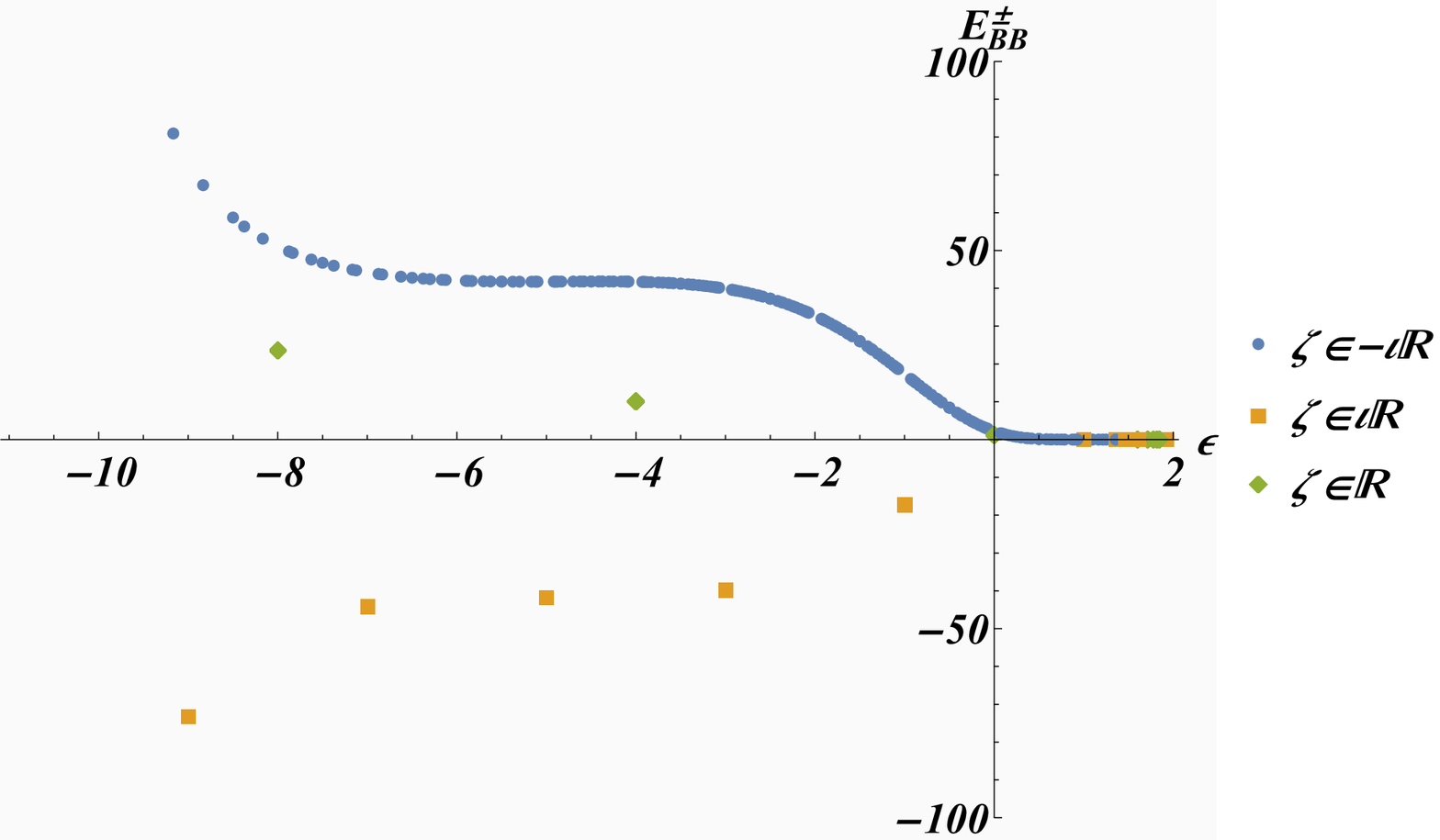, width=7.2cm} \epsfig{file=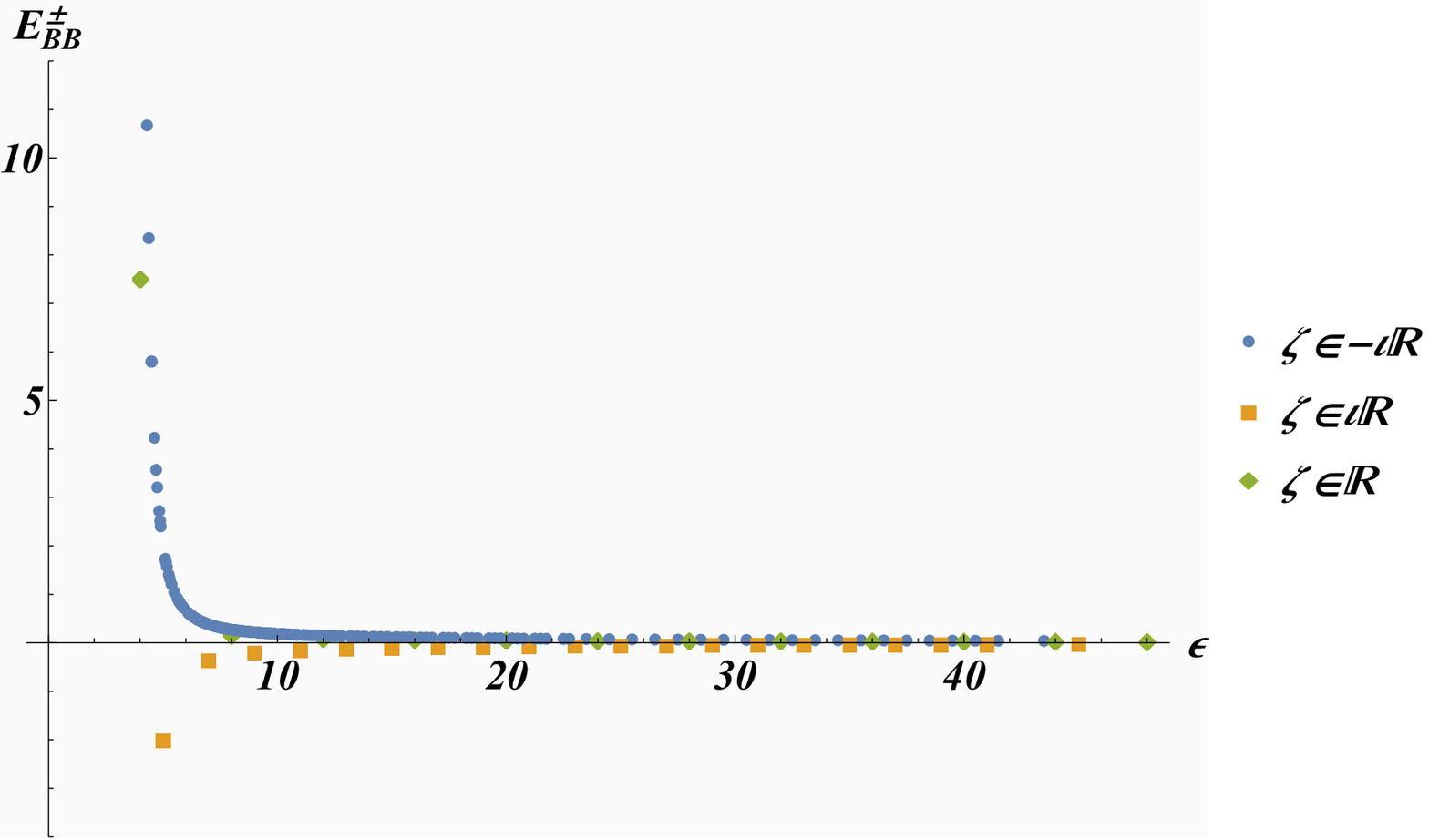, width=7.2cm}
\caption{Real energies $E_{\text{BB}}^{\pm }$ of the compacton (panel a) and unbounded (panel b) BPS solutions for 
the cases (\ref{c1}) - (\ref{c3}) with $\lambda =1$, $\mu =2$, $n=1$ and $c=0.2$ for several values of $\ell ,m$.  }
       \label{Fig1}}

We observe from figure \ref{Fig1} that the energies are finite and follow
distinct curves for the different cases. Moreover, for the case $\zeta \in
-\imath \mathbb{R}^{+}$ the curve is fairly dense and becomes more
connected\ when including more values for $\ell $ and $m$, hence $%
\varepsilon $. In the other cases this can not be achieved due to the
additional restriction on $\varepsilon $ so that the distribution is more
sparse. The transition at $\varepsilon =2$ is not smooth.

For these models the $\mathcal{CPT}^{\prime }$-symmetry identified from (\ref%
{cpt3}) must act as $\zeta \rightarrow -\zeta ^{\ast }$. For our solutions
in (\ref{sol}) this becomes $\zeta _{m}^{\pm }\rightarrow -(\zeta _{m}^{\pm
})^{\ast }=\zeta _{-m}^{\mp }$. Noting now that $\varepsilon (m,\ell
,s)=\varepsilon (-m,-\ell ,-s)$ in (\ref{c1}) and $\varepsilon (m,\ell
,s)=\varepsilon (-m,-\ell -1,-s)$ in (\ref{c2}), (\ref{c3}), we simply have
to choose a new $\ell ^{\prime }=-\ell $, $\ell ^{\prime }=-\ell -1$,
respectively, to obtain the same value for $\varepsilon $. This establishes
that $E[\zeta _{m}^{\pm }]=E[\zeta _{-m}^{\mp }]$ so that condition (iii) in
(\ref{cpt3}) also holds and the energy must therefore be real. Notice once
more that the $\mathcal{CPT}^{\prime }$-symmetry that ensures the reality of
the energies is different from $\mathcal{CPT}$, that was observed initially
for $V_{BB}(\zeta )$.

\section{Skyrme model with complex trigonometric potentials}

Next we study a model for which the Hamiltonian respects again the $\mathcal{%
CPT}$-symmetry: $\zeta \rightarrow -\zeta $, $\imath \rightarrow -\imath $,
but which has solutions transforming under a $\mathcal{CPT}^{\prime }$%
-symmetry to satisfy (\ref{cpt1}) with conditions (\ref{cpt2}) and/or (\ref%
{cpt3}) violated. Thus we are in the broken $\mathcal{CPT}^{\prime }$%
-regime. For this purpose we consider the variant of the model $\mathcal{L}%
_{0}+\mathcal{L}_{6}$ involving the trigonometric potential 
\begin{equation}
V_{T}(\zeta )=\sin ^{4}\zeta \cos ^{4}(\zeta +i\epsilon ),~~\ \varepsilon
\in \mathbb{R}.
\end{equation}%
We notice that unlike as in the pseudo Hermitian model discussed in section
3 only one of the factors in the potential is shifted so that the potential
is not simply boosted and most likely not pseudo Hermitian. The BPS
equations take the form 
\begin{equation}
\frac{n\lambda }{2\mu }\frac{\sin ^{2}\zeta }{\sqrt{V_{T}}}d\zeta =\pm
r^{2}dr\quad \Rightarrow \quad \frac{d\zeta }{dr}=\pm 3\alpha \cos
^{2}(\zeta +i\epsilon )r^{2},
\end{equation}%
where we abbreviated $\alpha :=\frac{2\mu }{3n\lambda }$. Integrating this
equation we find the solutions 
\begin{equation}
\zeta _{\alpha ,\gamma }^{\pm }(r)=-i\epsilon \pm \mathrm{arctan}\,\alpha
(r^{3}+\gamma )\,.
\end{equation}%
with integration constant $\gamma \in \mathbb{C}$. The symmetry identified
from condition (i) in (\ref{cpt1}) acts as $\mathcal{CPT}^{\prime }$: $\zeta
_{\alpha ,\gamma }^{\pm }\rightarrow -\left( \zeta _{\alpha ,\gamma }^{\pm
}\right) ^{\ast }=$ $\zeta _{\alpha ,\gamma }^{\mp }$. Thus the second
condition (\ref{cpt2}) still holds. However, the energies of the two
solutions related in this manner are in general not degenerate, i.e. $%
E[\zeta _{\alpha ,\gamma }^{+}]\neq E[\zeta _{\alpha ,\gamma }^{-}]$.
Depending on the nature of the integration constant $\gamma $ we find two
different types of behaviour and we can still find discrete values for the
two $\mathcal{CPT}^{\prime }$ related solutions that have degenerate
energies.

\subsubsection*{Real integration constants $\protect\gamma \in \mathbb{R}$}

Computing the energy $E_{\alpha ,\gamma }^{\pm }$ as in the previous
sections, the real and imaginary part acquire the form 
\begin{align}
\mathrm{Re}\,E_{\alpha ,\gamma }^{\pm }& =\frac{1}{32}\left( \frac{\pi }{%
\alpha }{-}\frac{2\gamma }{\alpha ^{2}\gamma ^{2}{+}1}\right) +\frac{1}{48}%
\left( \frac{2\gamma \left( \alpha ^{2}\gamma ^{2}{+}3\right) }{\left(
\alpha ^{2}\gamma ^{2}{+}1\right) ^{2}}{-}\frac{\pi }{\alpha }\right) \cosh
2\epsilon +\frac{\gamma \left( \alpha ^{2}\gamma ^{2}{-}3\right) \cosh
4\epsilon }{72\left( \alpha ^{2}\gamma ^{2}{+}1\right) ^{3}}  \notag
\label{eneim} \\
& +\frac{\gamma (2\cosh 2\epsilon {-}3)}{48\alpha }\mathrm{arctan}\,\alpha
\gamma \,, \\
\mathrm{Im}\,E_{\alpha ,\gamma }^{\pm }& =\mp \frac{\sinh 2\epsilon \left(
\lbrack 3\alpha ^{2}\gamma ^{2}{-}1]\cosh (2\epsilon ){+}3{+}3\alpha
^{2}\gamma ^{2}\right) }{36\alpha \left( \alpha ^{2}\gamma ^{2}{+}1\right)
^{3}}.
\end{align}%
This in general the energy is complex and we have $E_{\alpha ,\gamma
}^{+}=\left( E_{\alpha ,\gamma }^{-}\right) ^{\ast }$ and the model is in
the broken $\mathcal{CPT}^{\prime }$-phase. However, we note that the
imaginary part vanishes when parameterizing the integration constant as 
\begin{equation}
\gamma _{\ell }(\alpha ,\epsilon )=\ell \text{sech}\,\epsilon \frac{\sqrt{%
\cosh 2\epsilon {-}3}}{\sqrt{6}\alpha },\quad \quad \ell =\pm \,.
\label{gamma}
\end{equation}%
In this case we have also satisfied condition (iii) in (\ref{cpt3}) with $%
E[\zeta _{\alpha ,\gamma }^{+}]=E[\zeta _{\alpha ,\gamma }^{-}]$ and the $%
\mathcal{CPT}^{\prime }$-symmetry is restored. In order to keep the
condition $\gamma \in \mathbb{R}$, we must restrict $\left\vert \epsilon
\right\vert \in \lbrack \frac{1}{2}\mathrm{arccosh}\,3,\infty )$.

\subsubsection*{Purely imaginary integration constants $\protect\gamma \in i%
\mathbb{R}$}

Taking now $\gamma $ to be purely imaginary the $\mathcal{CPT}^{\prime }$%
-symmetry acts as $\mathcal{CPT}^{\prime }$: $\zeta _{\alpha ,\gamma }^{\pm
}\rightarrow -\left( \zeta _{\alpha ,\gamma }^{\pm }\right) ^{\ast }=$ $%
\zeta _{\alpha ,-\gamma }^{\mp }$. The real and imaginary parts of the
energies become now 
\begin{align}
\mathrm{Re}\,E_{\alpha ,\gamma }^{\pm }& =\frac{\pi }{32\alpha }\left( 1-%
\frac{2}{3}\cosh 2\epsilon \right) , \\
\mathrm{Im}\,E_{\alpha ,\gamma }^{\pm }& =\pm \frac{\sinh 2\epsilon \left(
\lbrack 1{+}3\alpha ^{2}\gamma ^{2}]\cosh 2\epsilon {-}3{+}3\alpha
^{2}\gamma ^{2}\right) }{36\alpha \left( 1{-}\alpha ^{2}\gamma ^{2}\right)
^{3}}+\gamma \frac{(2\cosh 2\epsilon {-}3)}{48\alpha }\mathrm{arctan}%
\,\alpha \gamma  \notag \\
& -\frac{\gamma }{72}\frac{\left( \alpha ^{2}\gamma ^{2}{+}3\right) \cosh
4\epsilon +\left( 1{-}\alpha ^{2}\gamma ^{2}\right) \left( 3{-}\alpha
^{2}\gamma ^{2}\right) \cosh 2\epsilon }{(1{-}\alpha ^{2}\gamma ^{2})^{3}}.
\label{Im2}
\end{align}%
Interestingly the real part becomes very simple and does not depend on the
integration constant $\gamma $. We may, however, still find values for $%
\gamma $ as function of $\alpha $ and $\epsilon $ for which the imaginary
part (\ref{Im2}) vanishes, but not in a closed form as in (\ref{gamma}). In
this case condition (iii) in (\ref{cpt3}) becomes $E[\zeta _{\alpha ,\gamma
}^{+}]=E[\zeta _{\alpha ,-\gamma }^{-}]$ and the $\mathcal{CPT}^{\prime }$%
-symmetry is also restored.

\section{A new Skyrme submodel with complex BPS solutions and real energy}

By decomposing the sigma model and the Skyrme term, Adam, Sanchez-Guillen
and Wereszczynski noticed in \cite{adam2017bps} that one may define further
consistent and solvable submodels by combining terms from either
decomposition as 
\begin{equation*}
\mathcal{L}_{+}^{(1)}:=\mathcal{L}_{2}^{(1)}+\mathcal{L}_{4}^{(1)},~~~\ ~%
\text{and~~~~~}\mathcal{L}_{+}^{(2)}:=\mathcal{L}_{2}^{(2)}+\mathcal{L}%
_{4}^{(2)}.
\end{equation*}%
Choosing the coupling constant in front of $\mathcal{L}_{4}$ to be negative
relative to $\mathcal{L}_{2}$, we consider now a slight modification of the
second submodel defined by the Lagrangian densities 
\begin{equation}
\mathcal{L}_{-}^{(2)}:=\lambda \left( \mathcal{L}_{2}^{(2)}-\mathcal{L}%
_{4}^{(2)}\right) ,~~~\ \ ~\ \ \ \lambda \in \mathbb{C}.
\end{equation}%
The corresponding Hamiltonian density for static solutions may then be
written as 
\begin{equation}
\mathcal{H}_{-}^{(2)}=\lambda (\bigtriangledown \zeta )^{2}-\lambda \sin
^{4}\zeta \sin ^{2}\Theta \left( \bigtriangledown \Theta \times
\bigtriangledown \Phi \right) ^{2}=A^{2}+\tilde{A}^{2},  \label{Hm}
\end{equation}%
where the dual fields are defined as%
\begin{equation}
A_{i}=\sqrt{\lambda }\zeta _{i},\qquad \text{and\qquad }\tilde{A}_{i}=\imath 
\sqrt{\lambda }\sin ^{2}\zeta \sin \Theta \varepsilon _{ijk}\Theta _{j}\Phi
_{k}.
\end{equation}%
Thus, the Hamiltonian density is of the same generic form as for the class
of general BPS models discussed in \cite{adam2013some}. Hence, following the
same reasoning, the imposition of a self-duality and anti-self-duality
between $A_{i}$ and $\tilde{A}_{i}$, 
\begin{equation}
A_{i}=\pm \tilde{A}_{i}  \label{AA}
\end{equation}%
selects out the BPS equations \cite{bogomol1976stability,prasad1975exact}.
Thus the energy functional $E_{-}^{(2)}$ for the solutions of (\ref{AA})
therefore acquires the form as in equation (\ref{Eb}).

We now solve the BPS equations (\ref{AA}) and subsequently compute the
energies $E_{-}^{(2)}$ for the solutions obtained. Multiplying (\ref{AA}) by 
$\Theta _{i}$, $\Phi _{i}\,$, $\zeta _{i}$ and summing over $i$ we obtain
the respective equations%
\begin{equation}
\zeta _{i}\Theta _{i}=0,~~\ \zeta _{i}\Phi _{i}=0,~~~\text{and~~~}\zeta
_{i}\zeta _{i}=\pm \imath \sin ^{2}\zeta \sin \Theta \varepsilon _{ijk}\zeta
_{i}\Theta _{j}\Phi _{k}.  \label{3}
\end{equation}%
The first two constraints are satisfied by a suitable choice of the
space-time dependence of $\Theta $, $\Phi \,$, $\zeta $. Since $\varepsilon
_{ijk}\zeta _{i}\Theta _{j}\Phi _{k}$ is simply the Jacobian for the
variable transformation $(x,y,z)\rightarrow (\Theta ,\Phi \,,\zeta )$, the
multiplication of\ the last equation by the volume element $d^{3}x$ in (\ref%
{3}) leads to%
\begin{equation}
(\bigtriangledown \zeta )^{2}d^{3}x=\pm \imath \sin ^{2}\zeta \sin \Theta
d\Theta d\Phi \,d\zeta .
\end{equation}%
Similarly as above, we choose spherical space-time coordinates $%
(x,y,z)\rightarrow (r,\theta ,\phi )$ with $r\in \lbrack 0,\infty )$, $%
\theta \in \lbrack 0,\pi )$, $\phi \in \lbrack 0,2\pi )$, identify $\Theta
=\theta $, $\Phi =n\phi $ with $n\in \mathbb{Z}$ and assume $\zeta (r)\in 
\mathbb{C}$. These choices will automatically solve the first two equations
in (\ref{3}), whereas the last one reduces to%
\begin{equation}
\frac{d\zeta }{dr}=\pm \imath \frac{n}{r^{2}}\sin ^{2}\zeta .  \label{diff}
\end{equation}%
Apart from the $\imath $, this equation coincides with equation (3.6) in 
\cite{adam2017bps} derived for $\mathcal{L}_{+}^{(2)}$ by expressing the
unit vector $\vec{n}$ by means of a stereographic projection. We solve
equation (\ref{diff}) to 
\begin{equation}
\zeta _{\pm }^{(m)}(r)=\imath \limfunc{arccoth}\left( c\mp \frac{n}{r}%
\right) +m\pi ,~~~\ \ ~\ \ \ c\in \mathbb{C}\text{,}m\in \mathbb{Z}.
\label{comp}
\end{equation}%
As seen in figure \ref{Fig2} the imaginary parts of these solutions tend to
zero for $r\rightarrow \infty $, whereas the real parts approach
asymptotically the constant value $m\pi +\tilde{c}/2$ when taking $c=\imath
\cot (\tilde{c}/2)$, $\tilde{c}\in \mathbb{R}\backslash \{2n\pi \}$ with $%
n\in \mathbb{Z}$. Moreover $\lim_{r\rightarrow 0}\zeta _{\pm }^{(m)}(r)=m\pi 
$.

\FIGURE{ \epsfig{file=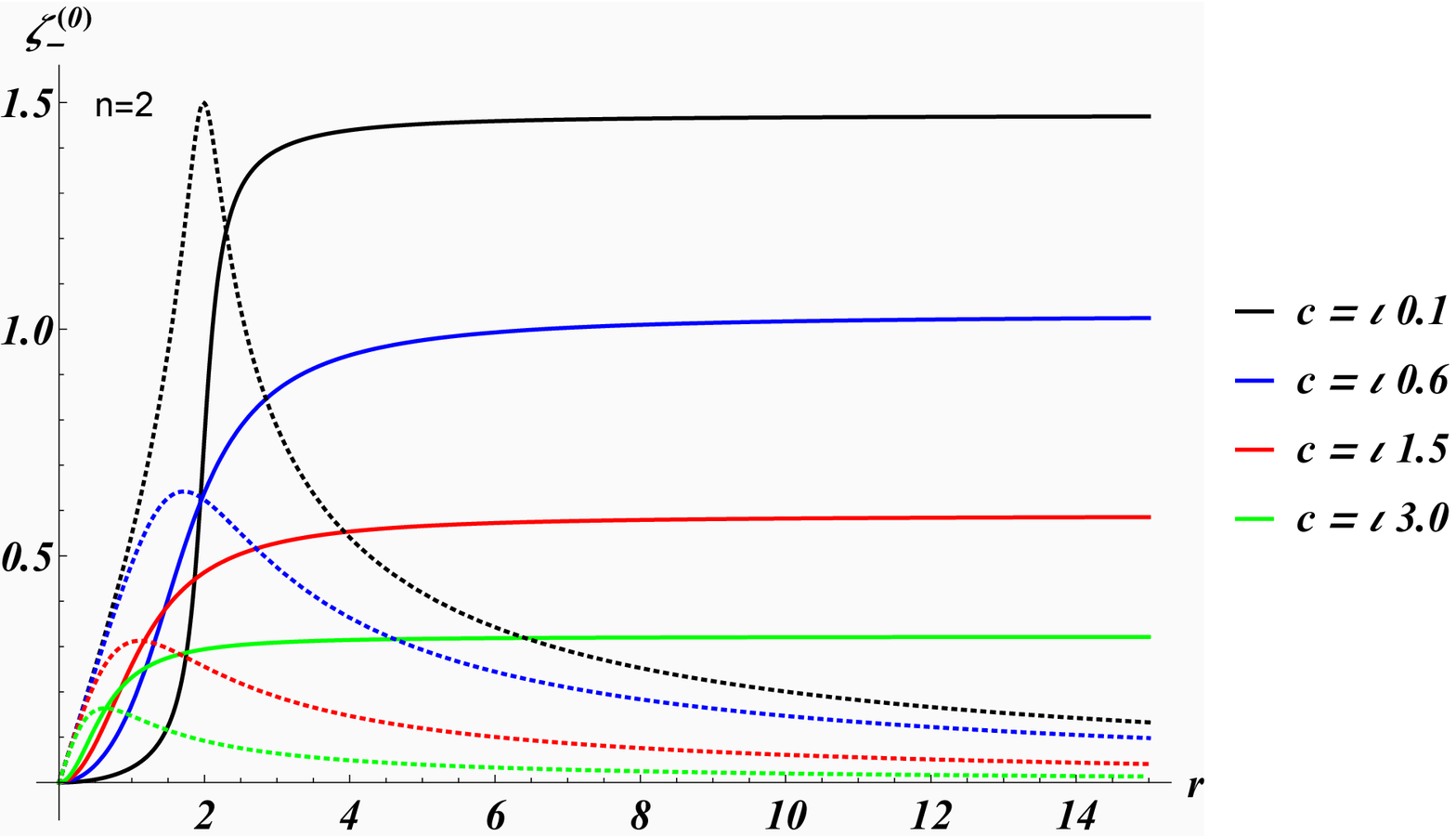, width=7.2cm} \epsfig{file=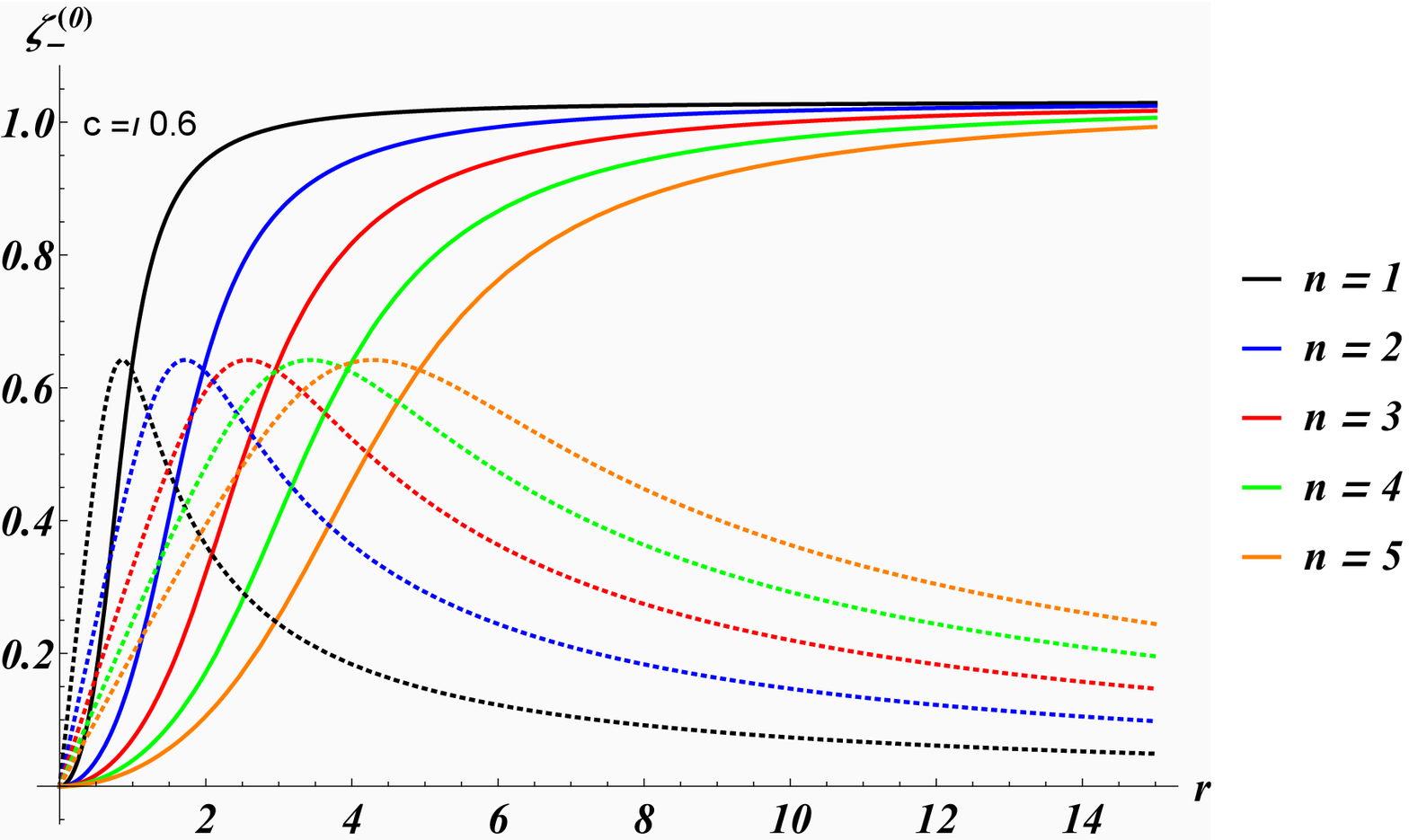,width=7.2cm}
\caption{Complex BPS solutions $\zeta _{- }^{(0)}$ for different values of $n$ and the initial condition $c$ for $\mathcal{L}_{-}^{(2)}$. Real parts as solid and imginary parts as dotted lines.}
       \label{Fig2}}

At first sight the solution (\ref{comp}) may seem to be unattractive due to
its complex nature. However, first of all it is continuous throughout the
entire range of $r$ and thus overcomes an issue of the real solutions $\zeta
_{r}^{(m)}=\func{arccot}\left( c-\frac{n}{r}\right) +m\pi $ found for $%
\mathcal{L}_{+}^{(2)}$ in \cite{adam2017bps}, which are discontinuous at $r=$
$n/c$. Moreover the energies for these solutions are real. We compute%
\begin{eqnarray}
E_{-}^{(2)}\left( \zeta \right) &=&\pm 2\imath \lambda \int d^{3}x\sin
^{2}\zeta \sin \Theta \varepsilon _{ijk}\zeta _{i}\Theta _{j}\Phi _{k} \\
&=&\pm 2\imath n\lambda \int \sin ^{2}\zeta \sin \theta d\theta d\phi
\,d\zeta  \notag \\
&=&\pm 8\pi \imath n\lambda \int_{0}^{\infty }\sin ^{2}\zeta \frac{d\zeta }{%
dr}dr  \notag \\
&=&\pm 8\pi \imath n\lambda \int_{\zeta (0)}^{\zeta (\infty )}\sin ^{2}\zeta
d\zeta =\left. \pm 2\pi \imath n\lambda \left[ 2\zeta -\sin (2\zeta )\right]
\right\vert _{\zeta (0)}^{\zeta (\infty )}.
\end{eqnarray}%
Thus taking now the complex coupling constant to be of the form $\lambda
=\imath \tilde{\lambda}$, $\tilde{\lambda}\in \mathbb{R}$, we obtain for
solutions $\zeta _{\pm }^{(m)}$ the real energies%
\begin{equation}
E_{-}^{(2)}\left( \zeta _{\pm }^{(m)}\right) =\pm 2\pi n\tilde{\lambda}\left[
\sin (\tilde{c})-\tilde{c}\right] .
\end{equation}%
We identify the $\mathcal{CPT}$-symmetry from (\ref{Hm}) as $\mathcal{CPT}%
:\zeta \rightarrow \zeta ^{\ast }$, which for our solution (\ref{comp})
becomes $\zeta _{\pm }^{(m)}(\tilde{c})\rightarrow \left[ \zeta _{\pm
}^{(m)}(\tilde{c})\right] ^{\ast }=$ $\zeta _{\mp }^{(-m)}(-\tilde{c})$. \
Since $E_{-}^{(2)}\left[ \zeta _{\pm }^{(m)}(\tilde{c})\right] =E_{-}^{(2)}%
\left[ \zeta _{\mp }^{(-m)}(-\tilde{c})\right] ,$ the energies are
guaranteed to be real by the antilinear symmetry $\mathcal{CPT}$.

\section{Conclusions}

We have studied several variants of the Skyrme Lagrangian density $\mathcal{L%
}$ in equation (\ref{L}). Our main focus has been on finding complex
solutions to the self-dual and anti-self-dual versions of the BPS-equation
or equation of motion. Identifying the $\mathcal{CPT}$-symmetries for these
models from the requirement in (\ref{cpt1}) allowed us to check the
remaining conditions (ii), (iii) in (\ref{cpt2}), (\ref{cpt3}) for the
constructed solutions, which when satisfied ensures the reality of the
energy. The broken $\mathcal{CPT}$-regime was also investigated by providing
a sample model in section 7 that is not pseudo Hermitian possessing generic
solutions for which neither of the conditions (ii) or (iii) holds. However,
when taking the integration constant to be real or purely imaginary and
parameterizing it in terms of the coupling constants of the model the
reality of the energy could be restored, which is also reflected in the
restored $\mathcal{CPT}$-symmetry of the model.

We found a number of novel Skyrmion configurations. Besides the standard
compacton or fractional compacton shapes of the BPS-Skyrmions, we found new
types of configurations such as complex kink, anti-kink, semi-kink, massless
solutions and purely imaginary compactons.

\medskip

\noindent \textbf{Acknowledgments:} FC was partially supported by Fondecyt
grant 1171475.

\newif\ifabfull\abfulltrue


\end{document}